\documentclass{aa}

\usepackage{graphicx}
\usepackage{txfonts}
\usepackage{color}
\usepackage{url}
\usepackage[flushleft]{threeparttable}
\usepackage[colorlinks=true, allcolors=blue]{hyperref}
\usepackage{orcidlink}
\usepackage{adjustbox}
\usepackage{multicol}
\usepackage{multirow}
\usepackage{xspace}
\usepackage{lastpage}
\newcommand {\gx}{{GX~13$+$1}\xspace}
\newcommand {\ixpe}{{IXPE}\xspace}
\newcommand {\nustar}{\textit{NuSTAR}\xspace}

\newcommand {\xrism}{{XRISM}\xspace}

\newcommand {\swift}{{Swift}\xspace} 
\newcommand{\fluxcgs}{erg\,
s$^{-1}$\,cm$^{-2}$}
\newcommand{\lum}{erg\,s$^{-1}$}

\begin{document}

\title{Peering through the Dip: IXPE unveils the Extended Scattering Environment of GX~13+1}

\titlerunning{Peering through the Dip: IXPE unveils the Extended Scattering Environment of GX~13+1}

\author{
Alessandro Di~Marco\inst{\ref{in:INAF-IAPS}}\thanks{Corresponding author: alessandro.dimarco@inaf.it}\orcidlink{0000-0003-0331-3259} 
\and
Fabio La~Monaca\inst{\ref{in:UTU},\ref{in:INAF-IAPS}}\orcidlink{0000-0001-8916-4156}
\and
Luigi Stella\inst{\ref{in:OAR}}\orcidlink{0000-0002-0018-1687}
\and
Anagha P. Nitindala\inst{\ref{in:UTU}}\orcidlink{0009-0002-7109-0202}
\and
Alexandra Veledina\inst{\ref{in:UTU},\ref{in:nordita}}\orcidlink{0000-0002-5767-7253}
\and
Anna Bobrikova\inst{\ref{in:UTU}}\orcidlink{0009-0009-3183-9742}
\and
Eleonora Veronica Lai\inst{\ref{in:OAC}}\orcidlink{0000-0002-6421-2198}
\and
Alessandro Papitto\inst{\ref{in:OAR}}\orcidlink{0000-0001-6289-7413}
\and
Maura Pilia\inst{\ref{in:OAC}}\orcidlink{0000-0001-7397-8091}
\and
Juri Poutanen \inst{\ref{in:UTU}}\orcidlink{0000-0002-0983-0049}
\and
Daniele Rogantini\inst{\ref{in:UC}}\orcidlink{0000-0002-5359-9497}
\and
Matteo Bachetti\inst{\ref{in:OAC}}\orcidlink{0000-0002-4576-9337}
\and
Maria Cristina Baglio\inst{\ref{in:OAB}}\orcidlink{0000-0003-1285-4057}
\and
Caterina Ballocco\inst{\ref{in:OAR}, \ref{in:Sapienza}}\orcidlink{0009-0001-0155-7455}
\and
Ettore Del Monte\inst{\ref{in:INAF-IAPS}}\orcidlink{0000-0002-3013-6334}
\and
Wei Deng\inst{\ref{in:GMNU}}\orcidlink{0000-0002-9370-4079}
\and
Giulia Illiano\inst{\ref{in:ICE}, \ref{in:IEEC}, \ref{in:OAB}}\orcidlink{0000-0003-4795-7072}
\and
Vladislav Loktev\inst{\ref{in:new}}\orcidlink{0000-0001-6894-871X}
\and
Sara Motta\inst{\ref{in:OAB}}\orcidlink{0000-0002-6154-5843}
\and
Fei Xie\inst{\ref{in:guan}}\orcidlink{0000-0002-0105-5826}
}

\authorrunning{A. Di Marco et al.}

\institute{
    INAF--Istituto di Astrofisica e Planetologia Spaziali, Via del Fosso del Cavaliere 100, I-00133 Roma, Italy \label{in:INAF-IAPS}
        \and
    Department of Physics and Astronomy, FI-20014 University of Turku, Finland \label{in:UTU}
        \and
    INAF--Osservatorio Astronomico di Roma, Via Frascati 33, I-00078 Monte Porzio Catone (RM), Italy \label{in:OAR}
        \and
    Nordita, Stockholm University and KTH Royal Institute of Technology, Hannes Alfvens vag 12, SE-10691 Stockholm, Sweden \label{in:nordita}
        \and
    INAF--Osservatorio Astronomico di Cagliari, Via della Scienza 5, I-09047 Selargius (CA), Italy \label{in:OAC}
        \and
    Department of Astronomy and Astrophysics, University of Chicago, IL-60637 Chicago, USA \label{in:UC}
        \and
    INAF--Osservatorio Astronomico di Brera, Via Bianchi 46, I-23807 Merate (LC), Italy \label{in:OAB}
        \and
    Dipartimento di Fisica, Sapienza Università di Roma, Piazzale Aldo Moro 5, I-00185 Rome, Italy \label{in:Sapienza}
        \and
    College of Physics and Electronic Information Engineering, Guangxi Minzu Normal University, CN-532200 Chongzuo, China \label{in:GMNU}
        \and
    Institute of Space Sciences (ICE, CSIC), Campus UAB, Carrer de Can Magrans s/n, E-08193 Barcelona, Spain \label{in:ICE}
        \and 
    Institut d’Estudis Espacials de Catalunya (IEEC), E-08860 Castelldefels (Barcelona), Spain   \label{in:IEEC}
        \and
    School of Mathematics, Statistics, and Physics, Newcastle University, NE1 7RU Newcastle upon Tyne, United Kingdom \label{in:new}
        \and
    Guangxi Key Laboratory for Relativistic Astrophysics, School of Physical Science and Technology, Guangxi University, CN-530004 Nanning, China \label{in:guan}
}
        
\date{Received 28 July 2026; accepted ***}

\abstract{Neutron star low-mass X-ray binaries feature complex accretion geometries, often including an accretion disk corona or disk winds. Here, a study of the highly inclined dipping source GX~13+1, using coordinated observations from the IXPE, NuSTAR, and Swift-XRT, is presented; this is the first time that such a campaign was conducted to monitor its periodic dip. Our analysis reveals significant variations in polarimetric properties tracking the dip. At the center of the dip, the polarization degree reaches a maximum of ${\sim}$9\% (at more than $8\sigma$\,CL). This is accompanied by a highly significant polarization angle rotation of roughly 60\degr when passing from the dip to the subsequent off-dip state. Furthermore, the dip state exhibits energy-dependent polarization. By modeling the polarization during the periodic dip, we constrain the geometry of the scattering medium. In the scenario of an oblate extended ADC, an equatorial radius at least 1.5 times its polar radius is needed with $\tau{\sim}0.3$. Alternatively, the results can be modeled as scattering in a disk wind with a most probable electron density of $1.3\times10^{14}$\,cm$^{-3}$ and an opening angle near 40\degr, corresponding to an optical depth of ${\sim}0.2$. The obtained results highlight the crucial role of X-ray polarimetric data to unveil the geometry of the extended scattering environment surrounding the accretion disks, advancing our understanding of neutron star low-mass X-ray binaries.
}


\maketitle
%
%-------------------------------------------------------------------

\section{Introduction}\label{sec:intro}

Low-mass X-ray binaries (LMXBs) hosting neutron stars (NSs) are crucial for understanding the properties of these extreme compact objects and the underlying causes of X-ray emission generated through mass accretion. NS-LMXBs exhibit a rich phenomenology, with their spectral and timing properties allowing them to be classified as Z and atoll sources according to their X-ray emission properties \citep{vanderklis89, hasinger89}. In NS-LMXBs, a compact Comptonizing region is present between the inner accretion disk and the NS surface; here the accreting matter decelerates before impacting the NS surface, forming a boundary layer (BL) around the equator of the NS \citep{Shakura88,Popham01} that may extend from the disk plane towards the poles of the NS, spreading over its surface, and forming the so-called spreading layer \citep[SL;][]{inogamov1999,suleimanov2006}. Alongside this compact Comptonizing region, certain systems exhibit phenomenology that can be explained by the presence of a hot, extended accretion disk corona \citep[ADC;][]{White1982, Miller2000,Psaradaki2018}. Such an ADC is expected to be optically thin in X-rays and inhabited by high-energy electrons, although its density, clumpiness, temperature, and geometry remain poorly constrained \citep{White1982, Miller2000}. Moreover, strong, non-collimated outflows of matter, referred to as disk winds \citep[DW; see][for a review]{wind2026}, are expected to occur at high accretion rates (approaching the Eddington luminosity). The presence of these disk winds is revealed by absorption line profiles as a major spectral feature, and sources at high inclination, such as \gx, are a good case study also for this component. 

\gx is a persistent NS-LMXB located at a distance of $7\pm1$ kpc in a binary system and has a late-evolved K5 III giant as a companion \citep{Bandyo99}. The soft-color/hard-color diagram (CCD) and hardness-intensity diagram (HID) of \gx are peculiar compared to other Z and/or atoll sources, challenging the determination of its state. Because of this, it was classified in the past as an atoll source \citep{hasinger89} with very high luminosity \citep[ranging from 0.5 up to 1.8 $L_\textrm{Edd}$;][]{Garcia88,XRISM2025}, which occasionally exhibits Z-like behavior, as seen in the secular evolution of its CCDs and HIDs \citep{Fridriksson_2015}. Furthermore, \gx exhibits a periodic dip every 24.5 days \citep{Iaria2014}, compatible with the orbital period of the source reported by \cite{Corbet2010} on the basis of the modulation of the K-band and X-ray light curves. The study of X-ray spectra obtained from different instruments indicated significant absorption along the line of sight, obscuring up to 80\% of the total emission in the energy band 0.6--2\,keV \citep{DT12}. This phenomenon can be explained by the existence of a DW \citep{XRISM2025,Neilsen2026,Rogantini2025} and/or an ADC \citep{Dai14,DiMarco25b}, suggesting a system inclination of 60–80$^\circ$. 

The Imaging X-ray Polarimetry Explorer \citep[\ixpe;][]{Soffitta2021, Weisskopf2022} has provided new insights into the geometry of NS-LMXBs; in particular, the peculiar polarimetric signatures can be used to probe the presence of disk winds \citep{Tomaru2024,nitindala25} or an extended accretion disk corona \citep{DiMarco25b,TomaruADC}. \ixpe has conducted three observations of \gx so far, reporting significant polarization in all of them and a peculiar variability in its polarimetric properties. A first observation, performed in October 2023, showed a polarization angle rotation of ${\sim}70$\degr \citep{Bobrikova24a}, and a dip also was detected in the light curve. No polarization angle variations were reported in the observation performed in February 2024, although indications for small PD variability were found \citep{Bobrikova24b}. Polarization variability associated with the presence of two dips in the light curve was reported for the April 2024 observation \citep{DiMarco25b}, with the degree of polarization (PD) reaching ${\sim}$4\% and the polarization angle (PA) exhibiting variations of ${\sim}$70\degr\ between the dip and off-dip states. The observed PA variability was ascribed to changes in the dominant emission component over time, under the condition that the axes of the NS and the accretion disk are not aligned \citep{Bobrikova24a,Bobrikova24b}. An alternative explanation is a change in obscuration along the line of sight that breaks the geometric symmetry of the X-ray emission due to scattering in an extended ADC or in the DW \citep{DiMarco25b}. A recent \xrism observation of \gx,  performed in February 2024, indicated the presence of a stratified wind with an unexpectedly slow and optically thick ($\tau{\sim}1$) component that reduces the overall X-ray emission \citep{XRISM2025} during a peculiarly obscured state (reported to occur in ${\sim}10$\% of the past \nustar and \textit{RXTE} observations) in which winds from \gx seem comparable to thermal-radiative winds, driven by X-ray irradiation of the outer disk. The study of the absorption lines also indicates the need for a scattered component that can be produced by scattering in the wind itself or by an extended accretion disk corona, in agreement with the interpretation reported by \citealt{DiMarco25b} for the \ixpe results. 

The present paper is based on coordinated \ixpe, \swift-XRT, and \nustar observations of \gx carried out in September 2025. Previous \ixpe observations detected erratic dips, while this new observation was performed for the first time during the periodic dip, when the region where the accretion stream impacts the outer region of the accretion disk causes a controlled increase in obscuration of the source. 
The data from X-ray observatories (unless specified otherwise) were extracted using the procedures reported in Appendix~\ref{sec:dataX}. In Sect.~\ref{sec:observations}, the observations are reported, and the polarimetric and spectral analyses are presented in Sect.~\ref{sec:data_results}. We discuss our results in Sect.~\ref{sec:discussion} and present our conclusions in Sect.~\ref{sec:conclusion}.

\section{Observations and state of the source}\label{sec:observations}

\begin{figure}[!t]
    \centering
    \includegraphics[width=\linewidth]{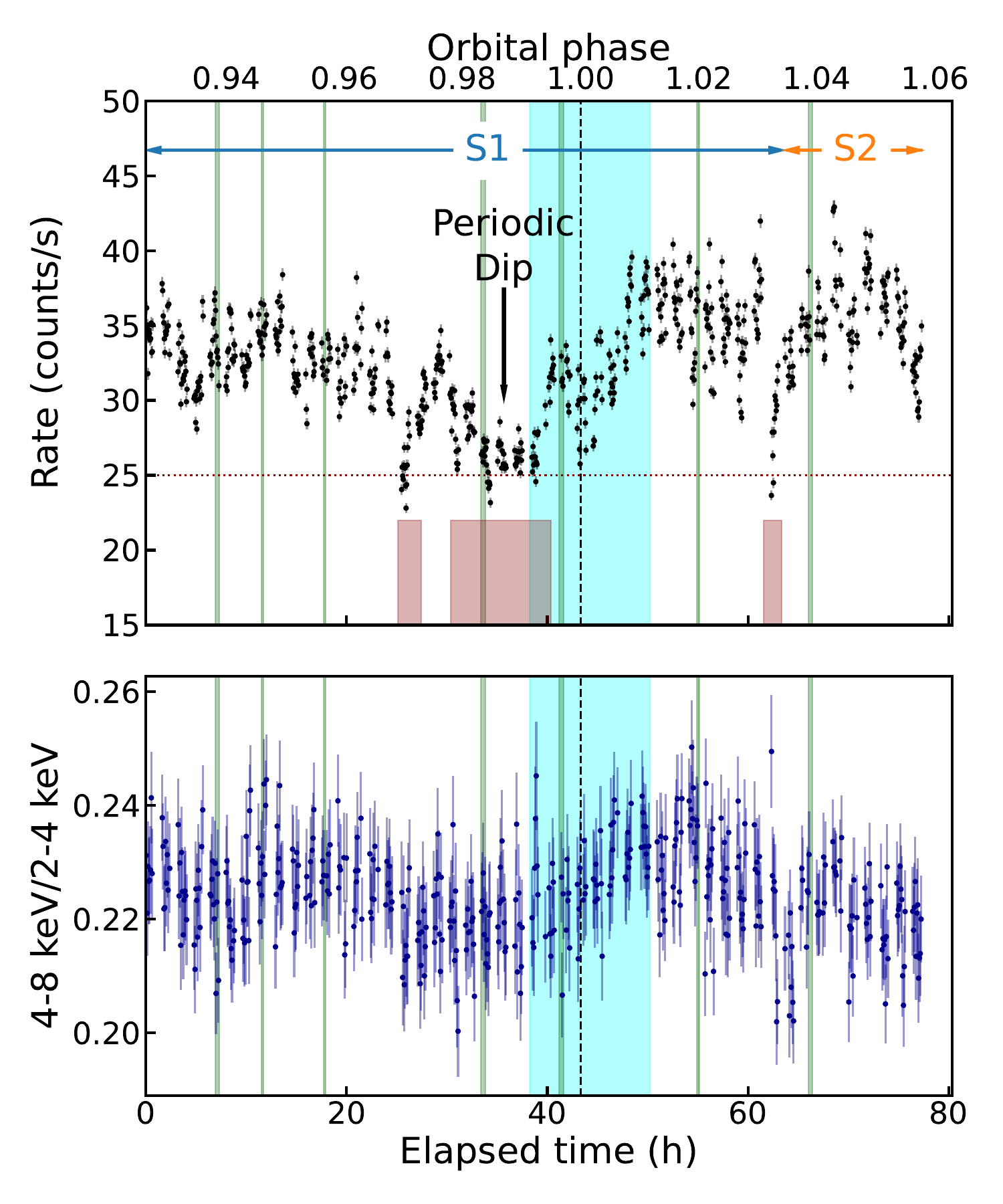}
    \caption{\ixpe light curve in 2--8\,keV (top) and the hardness ratio (bottom) for the present observation (time bins of 200\,s and 300\,s, respectively). The dashed vertical line reports the expected epoch of the dip, corresponding to orbital phase 1.0. The green and light blue bands report the periods of coverage by \swift-XRT and \nustar, respectively. The center of the observed periodic dip is indicated by an arrow, while the dark red bands identify the periods in which two narrow dips are observed (see text). The top horizontal arrows indicate two main segments chosen based on the change of the Stokes parameters (see Sect.~\ref{sec:data_results}).}
    \label{fig:ixpe_lc}
\end{figure}

This new \ixpe observation (ObsID 04002901) was performed for the first time during the periodic dip of \gx thanks to an approved \ixpe GO program (ID 2147, PI A.~Di~Marco). Figure~\ref{fig:ixpe_lc} shows the \ixpe light curve, with a vertical dashed black line reporting the expected epoch of the periodic dip, based on the ephemeris of \citealt{Iaria2014}. 
The \ixpe light curve shows complex flux variability. To identify the periodic dip, we selected time intervals in the data where the rate is less than 25\,cps, detecting three dips. The duration of each dip has been estimated as the time from the onset of the count-rate decrease to the end of the subsequent recovery. As reported in Figure~\ref{fig:ixpe_lc}, we detect a first dip at orbital phase ${\sim}0.970$ lasting ${\sim}2.5$\,h a second dip at orbital phase ${\sim}0.987$ lasting ${\sim}10$\,h, and a third dip for which we lost the ingress at orbital phase ${\sim}1.032$ lasting ${\sim}1.8$\,h. We identified the periodic dip as the one closest to orbital phase 1 and with the longest duration\footnote{\citealt{Iaria2014} report a periodic dip lasting ${\sim}10-30$\,h, depending on the instrument and the considered energy band}. This approach allowed us to identify the periodic dip, indicated by an arrow in Figure~\ref{fig:ixpe_lc}, which is at ${\sim}-0.19$\,days from the orbital phase 1, compatible with the jittering up to $0.15\pm0.15$\,d reported by \citealt{Iaria2014}.

The \ixpe observation was covered by a \nustar observation and several \swift-XRT pointings, as reported in Table~\ref{tab:exposure}.
\begin{table*}[!htb]
\centering
\caption{List of telescopes and/or instruments with the respective observation IDs, exposure times after regular data screening, and the covered orbital phase.}
\label{tab:exposure}
\begin{tabular}{llccccc}
\hline \hline
& Obs ID & Start (UTC) & Stop (UTC)  & Telescope & Exp. time (s) & Orb. phase \\
\hline
\ixpe & 04002901 & 2025-09-08 02:18 & 2025-09-11 07:38 & DU 1 & 147890 & 0.926--1.058 \\
 & & & & DU 2 & 148060 & \\
 & & & & DU 3 & 148090 & \\
\hline 
\swift-XRT  & 00036688054$^a$ & 2025-09-08 09:11 & 2025-09-08 14:01 & & 2248 & 0.938--0.946 \\
        & 00036688055 & 2025-09-08 20:02 & 2025-09-08 20:14 & & 635 & 0.956--0.957 \\
        & 00036688056 & 2025-09-09 11:41 & 2025-09-09 12:09 & & 1674 & 0.983--0.984 \\
        & 00036688057 & 2025-09-09 19:28 & 2025-09-09 19:55 & & 1644 & 0.996-0.997\\
        & 00036688058 & 2025-09-10 09:25 & 2025-09-10 09:49 & & 1474 & 1.019--1.020\\
        & 00036688059 & 2025-09-10 20:17 & 2025-09-10 20:42 & & 1554 & 1.038--1.039\\
\hline
\nustar & 91101332002 &2025-09-09 16:29 & 2025-09-10 04:32 & FPMA &  10579 & 0.991--1.012\\
 & & & & FPMB & 10740 & \\
\hline
\end{tabular}
\tablefoot{$^a$This \swift-XRT observation includes two pointings lasting ${\sim}1560$\,s and ${\sim}680$\,s, respectively, as reported in Fig.~\ref{fig:ixpe_lc}.
}
\end{table*}
The \nustar observation was performed around the expected epoch of the periodic dip, where one of the \swift-XRT observations was performed, while the other \swift-XRT observations were distributed throughout the \ixpe observation (see Fig.~\ref{fig:ixpe_lc}). 

\begin{figure}[!hbt]
    \centering
    \includegraphics[width=0.78\linewidth]{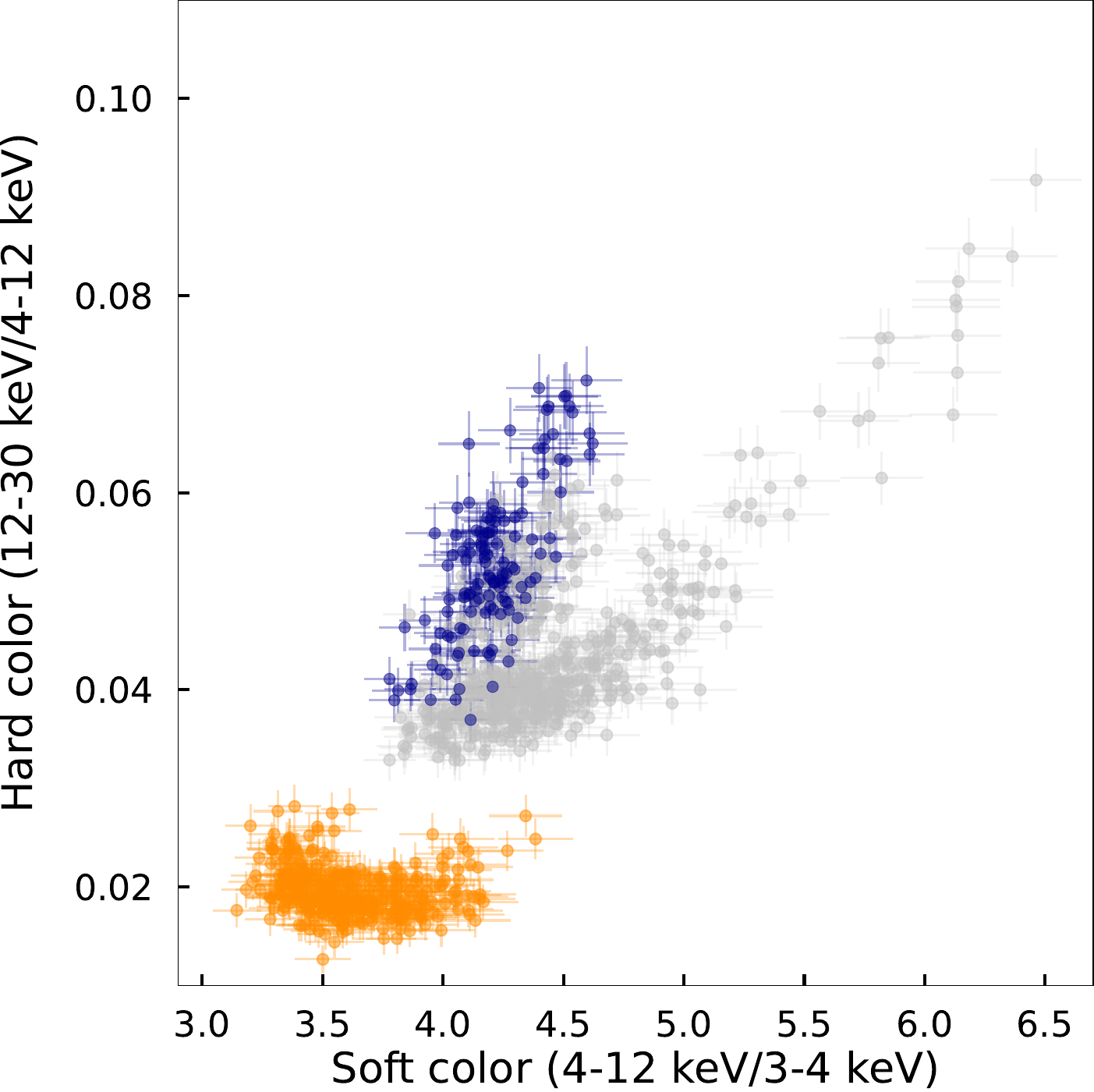}  \vspace{0.3cm}\\
    \includegraphics[width=0.78\linewidth]{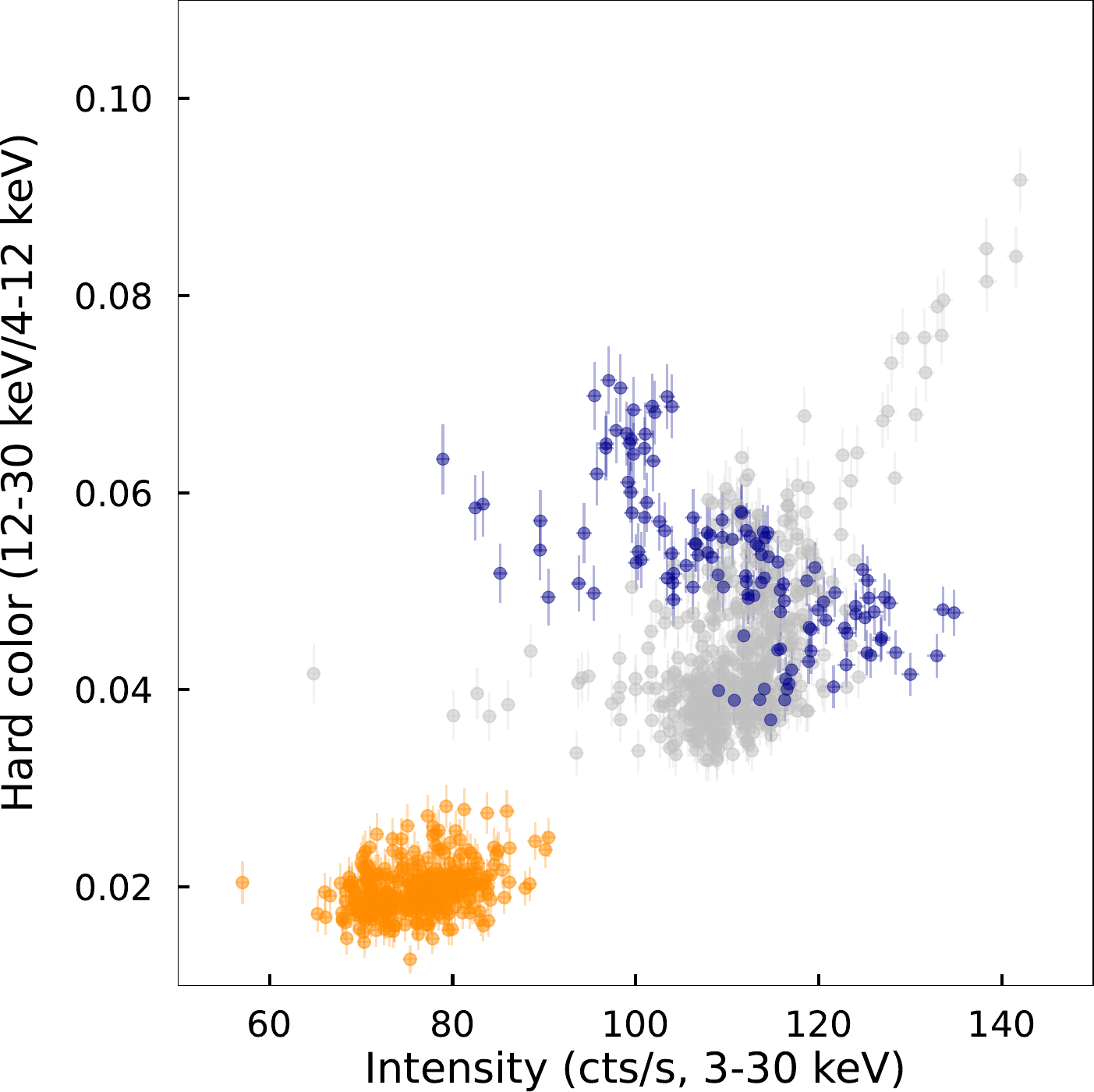}
    \caption{\nustar CCD (top) and HID (bottom) in 128\,s time bins. The blue points refer to the present observation; in orange are reported the points from observation ID 30901010002 corresponding to a peculiarly obscured state \citep{XRISM2025}; the grey points are from the other two \nustar observations of \gx available in the \nustar archive.}
    \label{fig:hid_ccd}
\end{figure}

To identify the state of \gx during this campaign, we used all the \nustar observations available on HEASARC to obtain the CCD and HID of Fig.~\ref{fig:hid_ccd}. The observation with ID 30901010002, reported in orange, was performed in February 2024 simultaneously with \ixpe and \xrism and corresponds to a peculiarly absorbed state \citep{XRISM2025}.
The \gx track in the \nustar CCD shows the pattern of a Z source, and the present observation was performed in the normal branch (NB), while the HID shows a peculiar pattern, as reported in the past \citep{Stella1985, Fridriksson_2015}, with the NB crossing the flaring branch (FB); the horizontal branch (HB) is not observed in these diagrams. 

\section{Data analysis and results}\label{sec:data_results}

The results presented in this section are based on the data handling reported in Appendix~\ref{sec:dataX}. As previously stated, the \ixpe light curve of Fig.~\ref{fig:ixpe_lc} shows three dips: the two narrow dips at orbital phases ${\sim}0.970$, and ${\sim}1.032$ (lasting ${\sim}2.5$, and ${\sim}1.8$ hours, respectively) and the main periodic dip at orbital phase ${\sim}0.987$ lasting ${\sim}10$\,h. The main dip is associated with the periodic passage of the bulge, which is the region where the disk became thicker due to the impact of the accretion stream. The flux during these dips reduces from an average of ${\sim}$38\,counts/s to less than ${\sim}25$\,counts/s; this reduction of ${\sim}35$\% is in agreement with the value observed during two erratic dips by \cite{DiMarco25b} and the dip observed by \ixpe in the first observation of \gx \citep{Bobrikova24a}. 

\subsection{Polarimetric analysis}

\begin{figure}[!ht]
    \centering
    \includegraphics[width=0.8\linewidth]{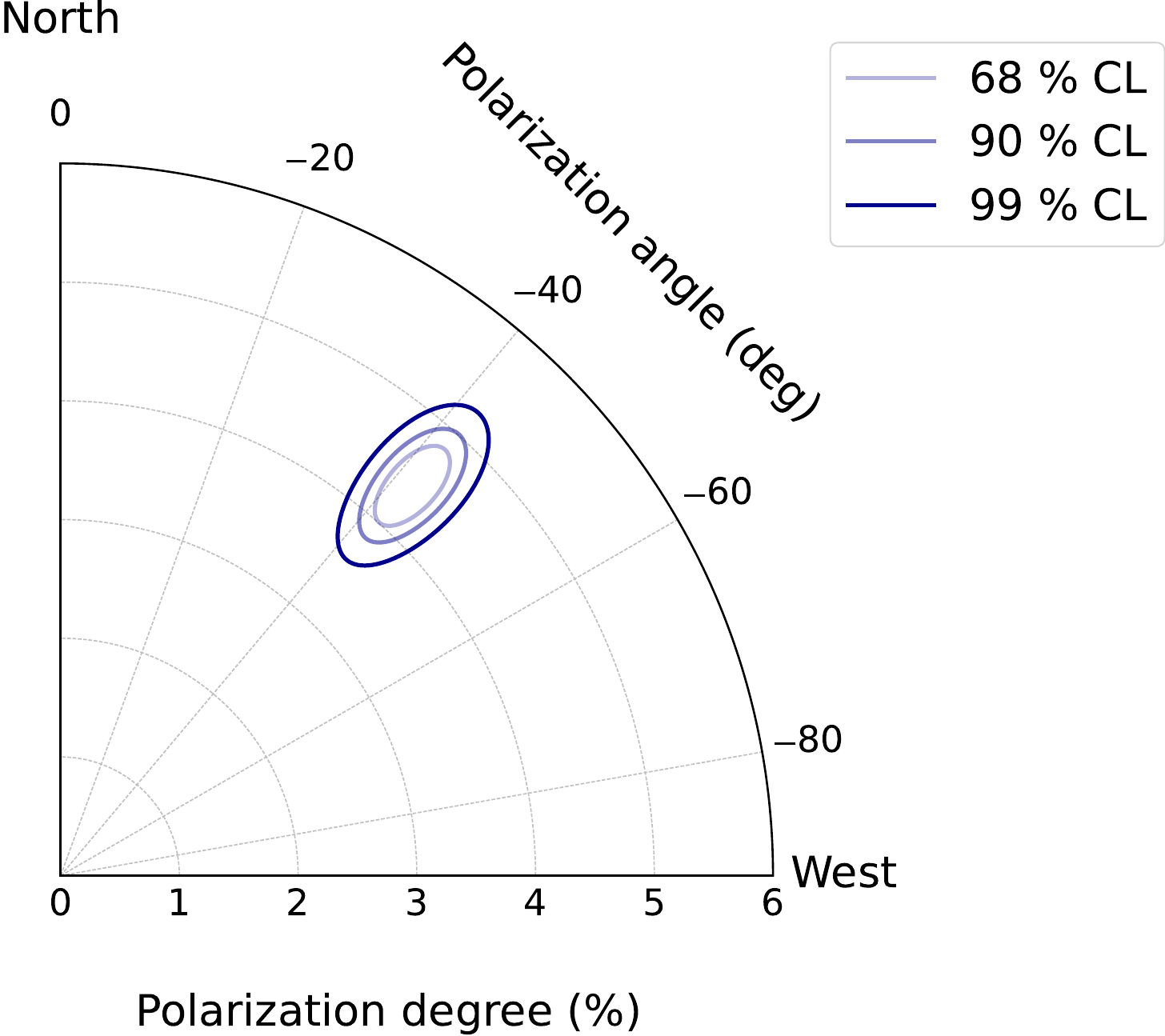}
    \caption{Allowed regions for PD and PA at different confidence levels.}
    \label{fig:pd_average}
\end{figure}

In this section, we report the model-independent analysis performed using the \texttt{pcube} algorithm in the \texttt{xpbin} tool of the \textsc{ixpeobssim} software \citep{Baldini22} applying the unweighted approach \citep{DiMarco_2022}. Considering the full new observation performed in September 2025, \ixpe data in the nominal 2--8\,keV energy band show a polarization of $\textrm{PD} = 4.4\% \pm 0.3\%$ with $\textrm{PA} = -42.0 \degr \pm 1.7\degr$ (hereafter, the errors are at the 68\% confidence level, CL, unless otherwise stated). The result is significant at $15\sigma$ CL. The corresponding protractor plot is in Fig.~\ref{fig:pd_average}. 

\begin{figure}[!htb]
    \centering
    \includegraphics[width=0.9\linewidth]{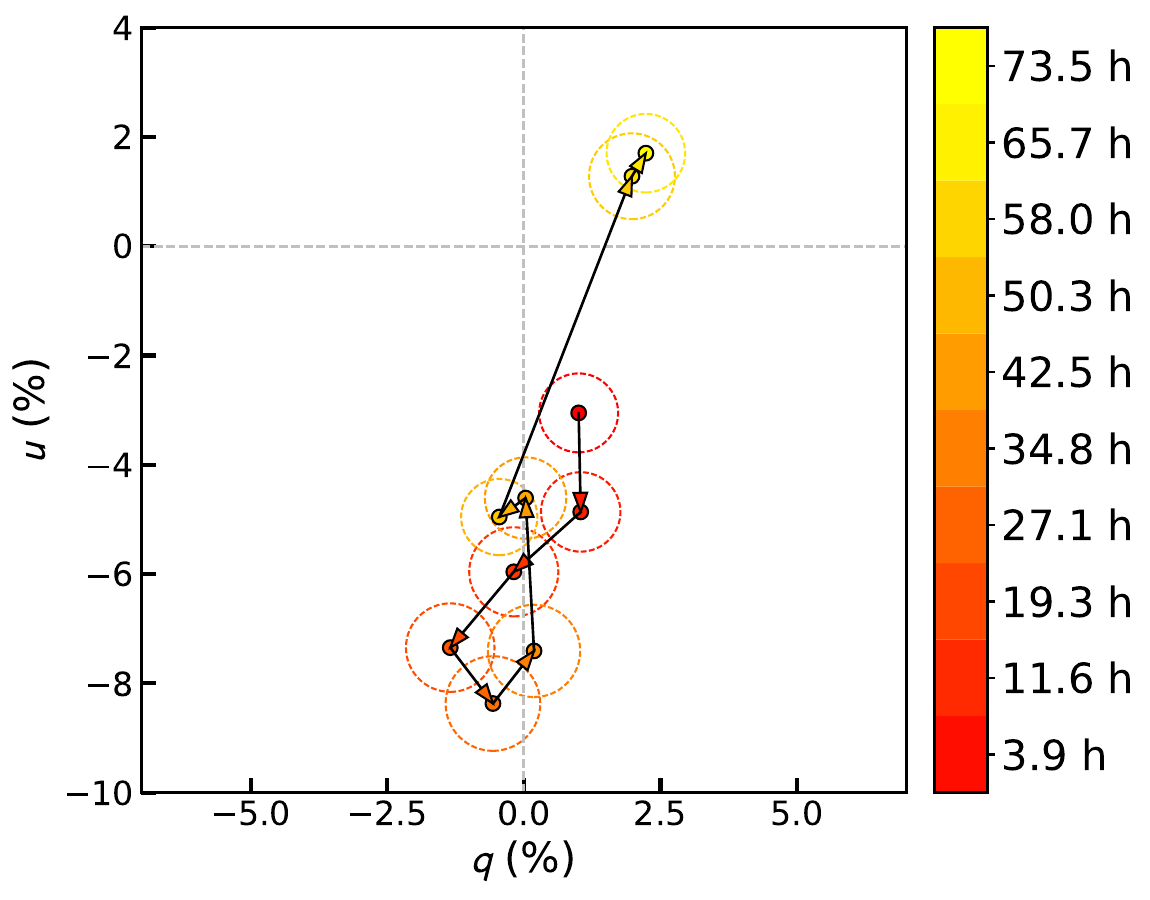}
    \caption{Time dependence of the normalized Stokes parameters $q$ and $u$ using 7.7\,h time bins. The colorbar and the arrows show the time evolution. Confidence regions are reported at 68\% CL.}
    \label{fig:qu_time}
\end{figure}

\begin{figure}[!t]
    \centering
    \includegraphics[width=0.85\linewidth]{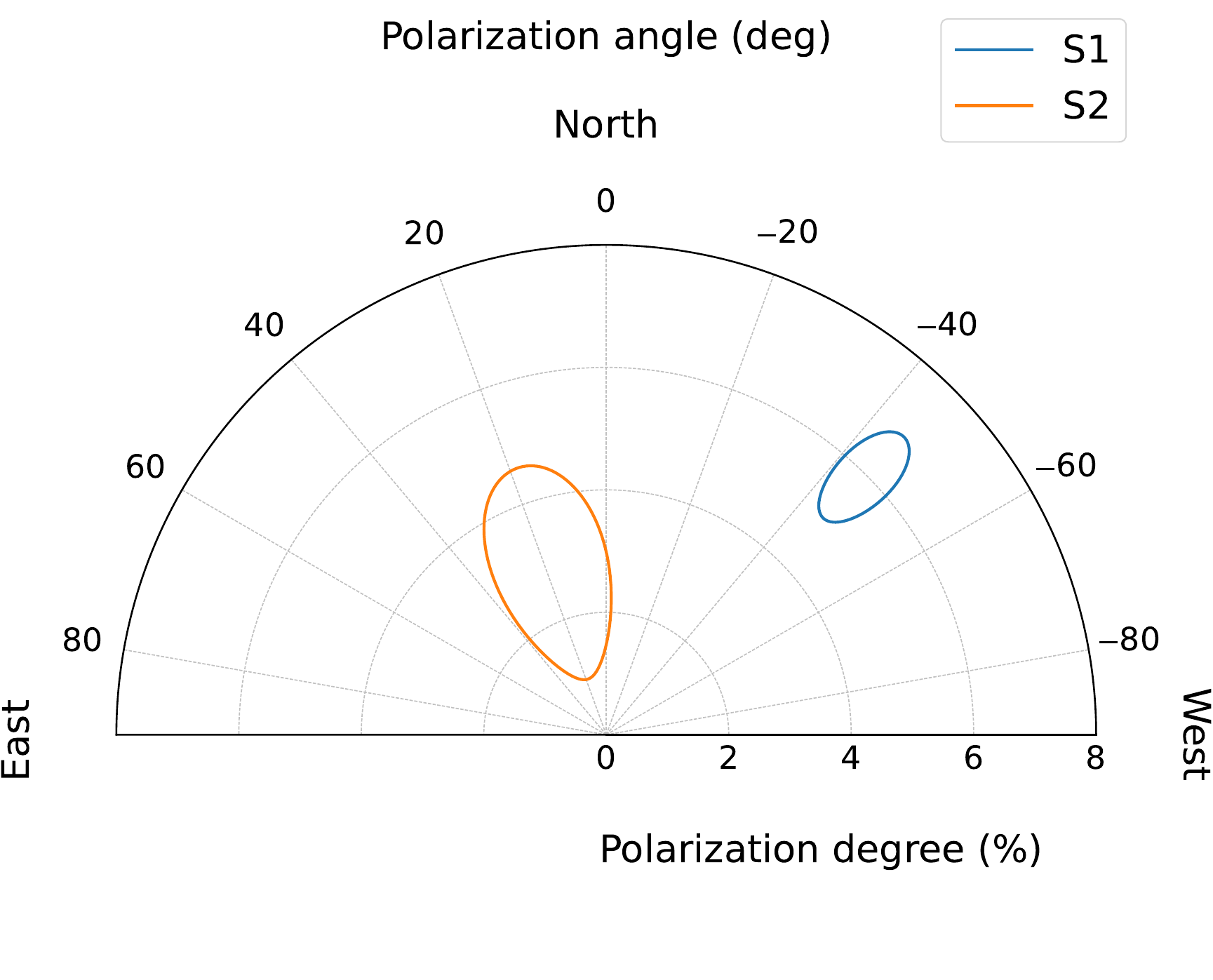}
    \caption{Allowed regions for PD and PA in the S1 (blue) and the S2 (orange) periods. Regions are reported at 99\%~CL.}
    \label{fig:pd_flux}
\end{figure}

To investigate possible temporal variability in the polarimetric properties along this new observation of \gx, we applied a constant ${\sim}8$\,h time binning, obtaining the polarization as a function of time reported in Fig.~\ref{fig:qu_time}; constant PD and PA are excluded by the $\chi^2$ test ($\chi^2$/dof=50/9, and $\chi^2$/dof=164/9, for PD and PA, respectively). 
The polarimetric pattern follows a trajectory tracking the presence of the main dip. In particular, at the beginning of the observation, the source is already in a dipping state with polarization moving toward a maximum of PD, which is reached in the center of the periodic dip; after the egress of the second narrow dip (orbital phase ${\sim}1.032$), the flux increases and the polarization changes, reaching a lower PD. This result shows that the observation can be divided into two main parts, as indicated in Fig.~\ref{fig:ixpe_lc}: Segment 1 (S1), whose polarization is dominated by the dip event, and Segment 2 (S2), which begins at orbital phase${\sim}1.034$ (after 63.6\,h since the beginning of the observation). The results obtained for these two main states of the observation are reported in Fig.~\ref{fig:pd_flux}, where PA changes significantly from $-45.0\degr \pm 1.5\degr$ (S1) to $18\degr \pm 6\degr$ (S2), while PD decreases from $5.9\%\pm0.3\%$ to $2.8\% \pm 0.6\%$; the total PA variation is ${\sim}60\degr$ passing from S1 to S2.

\begin{figure}[!tb]
    \centering
    \includegraphics[width=\linewidth]{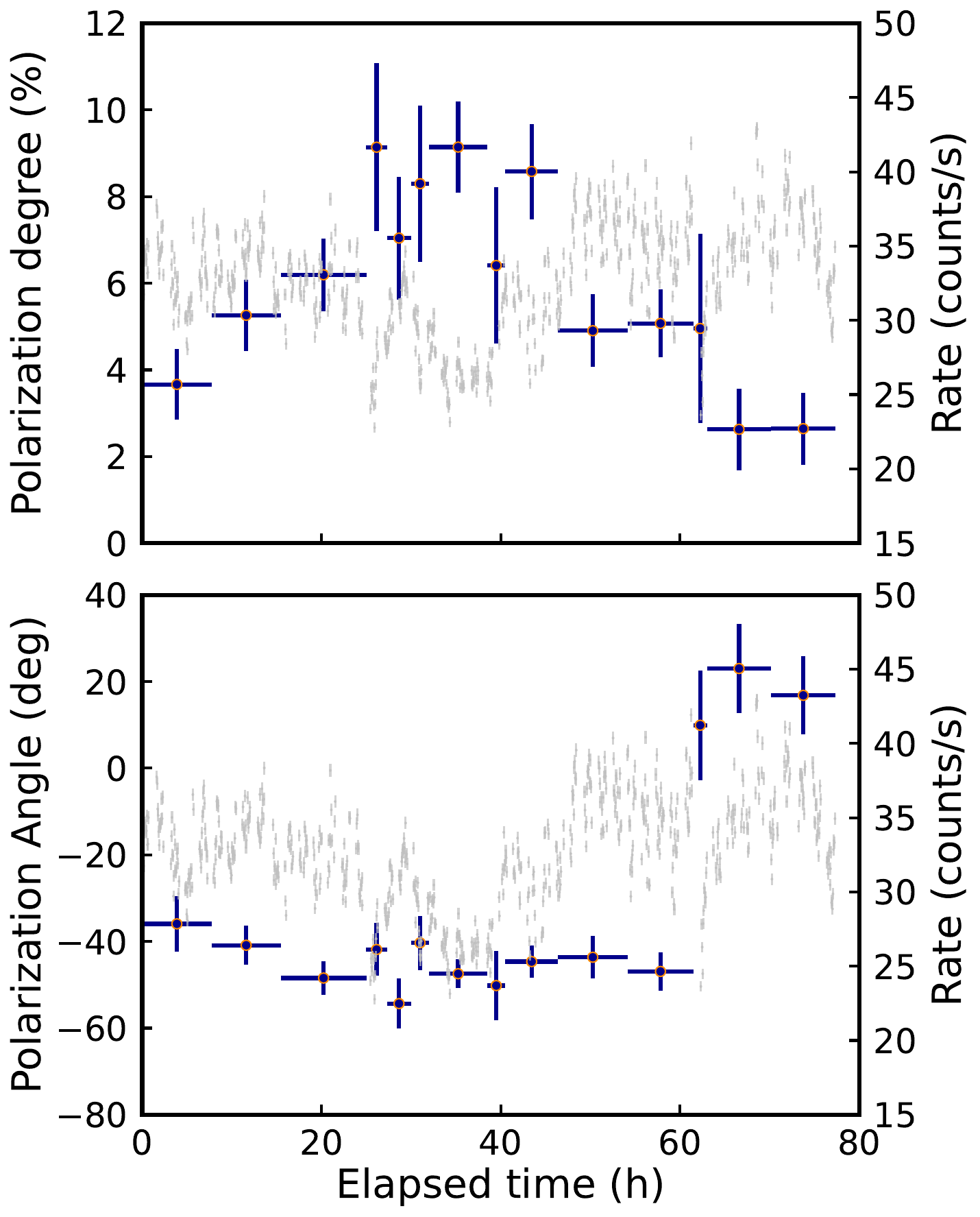}
    \caption{Polarization properties of \gx as a function of time with narrow selections of the dips. The panels from top to bottom show PD and PA, respectively. The gray points in both panels show the 2--8 keV count rate (right axis). Errors are at 68\% CL.}
    \label{fig:pd_time_dips}
\end{figure}

A deeper investigation of polarimetric time variability has been performed using narrow selections around the two narrow dips (indicated by red bars in Fig.~\ref{fig:ixpe_lc}) and by selecting the ingress and egress of the periodic dip; the results are reported in Fig.~\ref{fig:pd_time_dips}. The polarization degree and angle appear to be related to the source flux, with PD reaching the maximum during the first narrow dip and the periodic dip; the second narrow dip, for which we do not observe the ingress, has a lower significance, and both PD and PA seem to align more with S2 than S1. The polarization angle during the observation shows an indication for a rotation from north to west as the flux decreases and, during the second narrow dip, rotates toward the same PA as in S2. No significant difference in the polarization is observed during the entrance and the egress of the periodic dip in the present data, and the polarization in the middle of it reaches the maximum value of $9.1\% \pm 1.1\%$ at $-47\degr \pm 3\degr$ (significant at 8.3$\sigma$ CL).
These time-resolved analyses followed the procedures adopted in the previous studies of \gx \citep{Bobrikova24a, Bobrikova24b, DiMarco25b}, and the results are compatible with those observed by \cite{DiMarco25b}, where PD and PA appear to be related to the source flux. 

\begin{figure*}[!bt]
    \centering
    \includegraphics[width=0.48\linewidth]{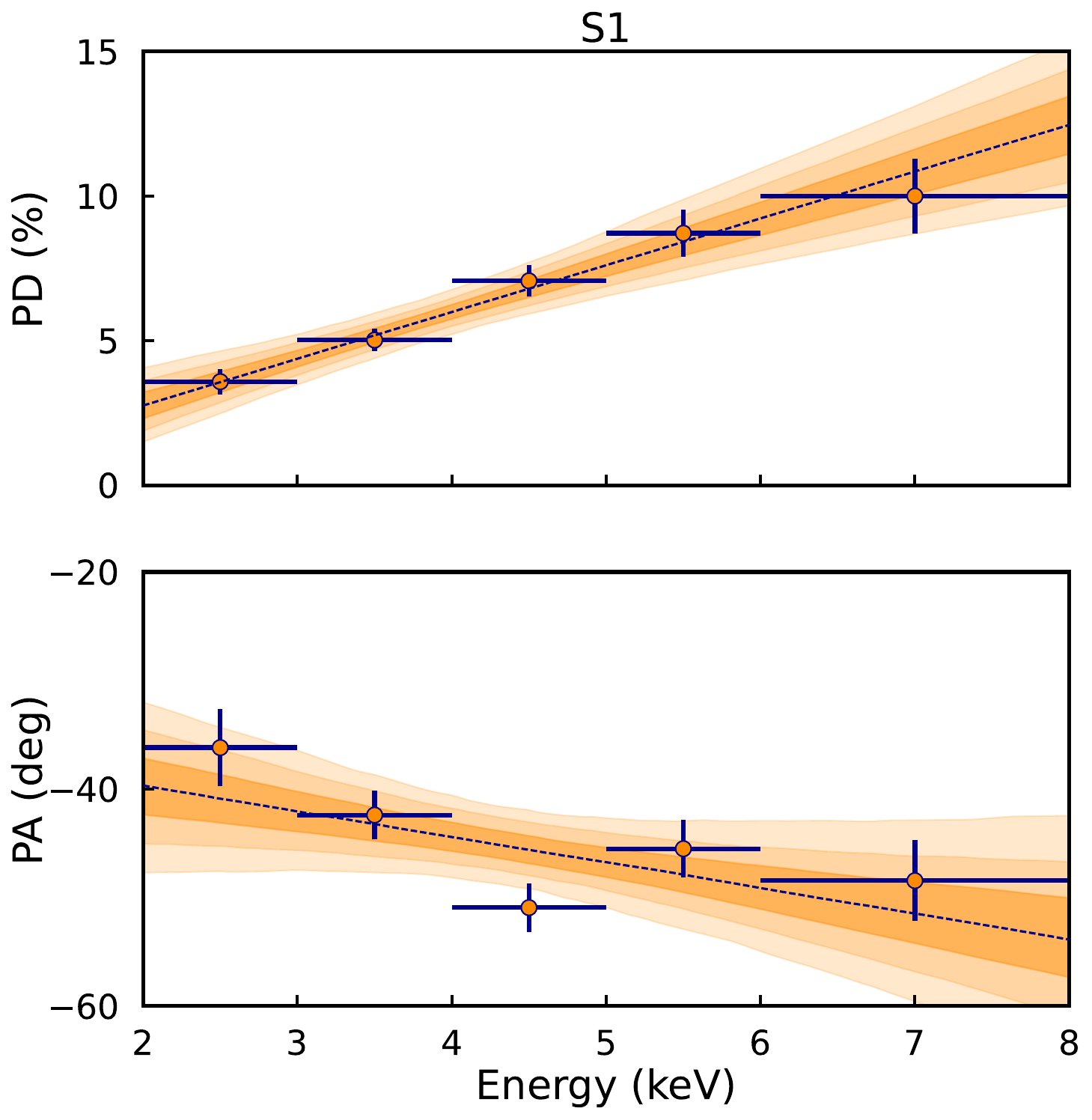}
    \includegraphics[width=0.48\linewidth]{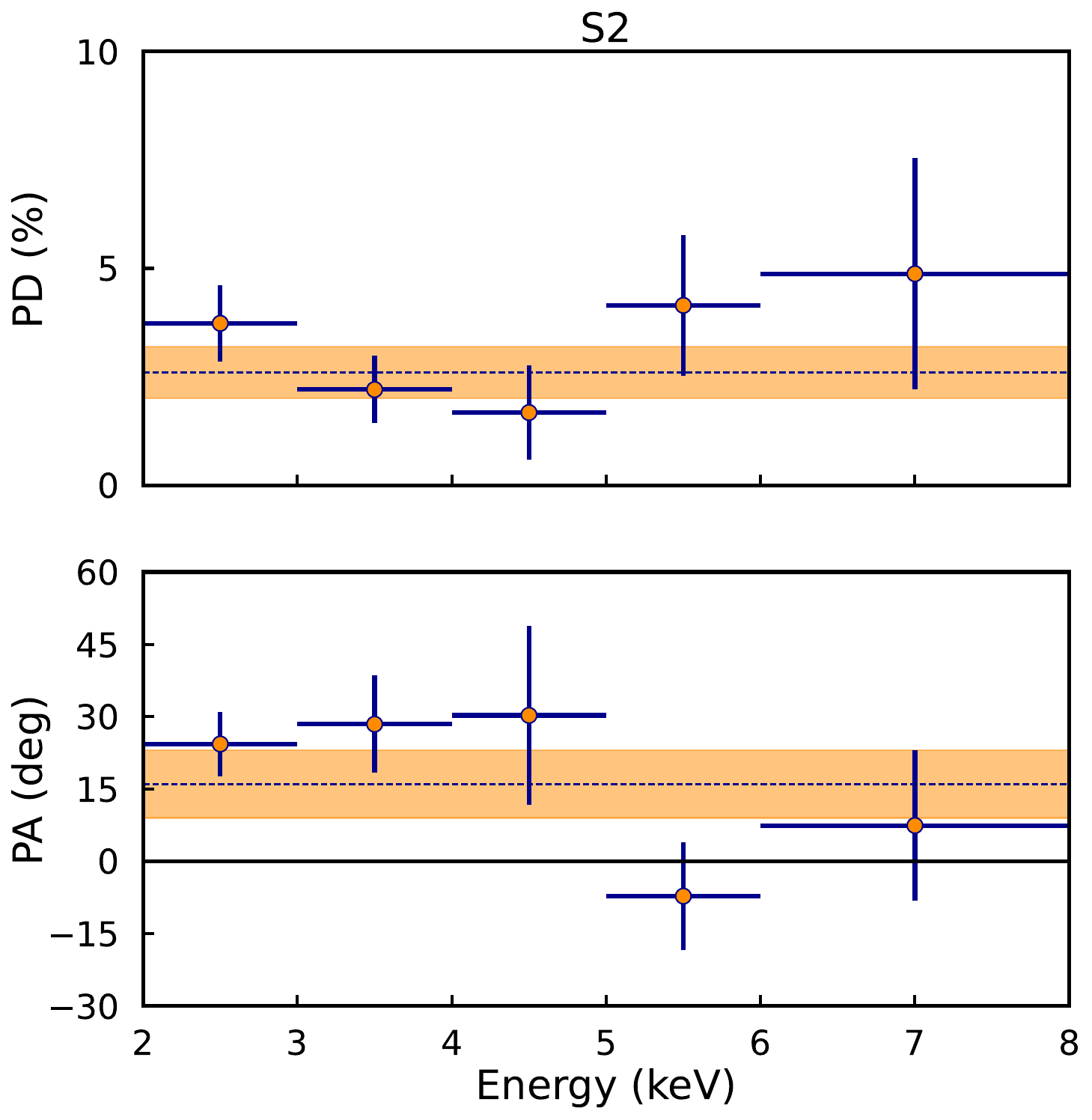}
    \caption{\gx polarization degree (top) and angle (bottom) as a function of energy for the S1 (left) and the S2 (right) segments. Error bars are at 68\% CL. In the left panels, the blue line represents the best fit for the linear trend, and the shaded regions indicate the uncertainties on it at $1\sigma$, $2\sigma$, and $3\sigma$~CL. In the right panels, the blue lines represent the average PD and PA as obtained in Fig.~\ref{fig:pd_flux}, with orange bars reporting 68\%~CL on these values.}
    \label{fig:pol_energy}   
\end{figure*}

An energy-resolved analysis was performed for both S1 and S2, as reported in Fig.~\ref{fig:pol_energy}.
A constant PD with energy is excluded in S1 ($\chi^2$/dof=60/4), while a PD increasing with energy is well described by a linear trend with an intercept of $(-0.5 \pm 0.8)\%$ and a slope of $(1.6\pm0.2)$\%/keV with $\chi^2$/dof=1.1/4. The resulting fit is reported in the top-left panel of Fig.~\ref{fig:pol_energy}, as obtained by \texttt{ultranest} \citep{ultranest}. Also, a constant PA with energy can be excluded in S1 at 99.6\%~CL ($\chi^2$/dof=15/4), while a fit with a linear trend provides an intercept $-35\degr\pm4\degr$ and a slope $-(2.3\pm0.9)$\degr/keV with $\chi^2/dof=9.1/4$; the poor $\chi^2$/dof obtained for the PA linear trend is mainly due to the point in the energy bin 4--5\,keV, as visible in the bottom-left panel of Fig.~\ref{fig:pol_energy}. For S2, both PD and PA are consistent with constant values, equal to the ones obtained in Fig.~\ref{fig:pd_flux} ($\chi^2$/dof=4.3/4, and $\chi^2$/dof=8/4, respectively). This change in the energy dependence of polarization across an erratic dip was also reported by \cite{Bobrikova24a}, although with a lower significance. The same study for energy dependence was performed in each time bin of Fig.~\ref{fig:pd_time_dips}, but the significance is too low to obtain any conclusion. For the periodic dip, a constant PD provides $\chi^2$/dof=7.4/4 (excluded at 88\% CL), while a $\chi^2$/dof=5/4 is obtained for a linear trend having an intercept of $(5 \pm 3)\%$ and a slope of $(1.0\pm0.7)$\%/keV; the constant PA provides $\chi^2$/dof=3.0/4, while the linear trend results in an intercept $-45\degr\pm8\degr$ with slope $-(1\pm2)$\degr/keV ($\chi^2/dof=2.5/4$). In conclusion, for the dip, a linear trend with energy for PD seems to be favored, although with a low significant slope, while a constant trend for the PA is sufficient to describe the data given the present uncertainties.

\subsection{Spectral analysis}

The \swift-XRT and \nustar observations (see Fig.~\ref{fig:ixpe_lc} and Table~\ref{tab:exposure}) allow for a model-dependent spectropolarimetric analysis thanks to proper broadband spectral modeling. Firstly, the \swift-XRT observations were analyzed by using the simple model \texttt{TBabs*(diskbb+bbodyrad)} \citep{Bobrikova24b,DiMarco25b} in \textsc{xspec} \citep{Arnaud96}. Hereafter, in the interstellar absorption model \texttt{tbabs}, we set the element abundances at the \texttt{wilm} values \citep{Wilms2000} and the photoelectric absorption cross-sections to \texttt{vern} \citep{Vern95}. A \texttt{gabs} component was included with energy frozen at 1.84\,keV to obtain a satisfactory fit; this could be a \swift-XRT instrumental effect due to the Si~K-edge \citep{swift_silicon} or the presence of silicates in the interstellar dust along the line of sight of \gx \citep{Rogantini2020, Rogantini2025, Vaia2026}. No emission lines are visible in the \swift-XRT data; the results of this spectral analysis are reported in Table~\ref{tab:spectrum1}, and show an average \texttt{diskbb} temperature at ${\sim}0.8$\,keV.
\begin{table*}[!htb]
\centering
\caption{Best-fit parameters for the spectral model \texttt{tbabs*(diskbb+bbodyrad)} as obtained from different \swift-XRT observations in the 0.8--10.0~keV energy band. Errors are reported at 68\%~CL.}
\label{tab:spectrum1}
\label{tab:spectrum} 
\begin{tabular}{crcccccc}
\hline \hline
 Observation ID & & 36688054 & 36688055 & 36688056 & 36688057 & 36688058 & 36688059 \\ \hline \\ [-1.7ex] 
\texttt{tbabs} & $N_{\text{H}}$ (10$^{22}$ cm$^{-2}$) & $4.79_{-0.17}^{+0.13}$ & $5.1\pm0.2$ & $4.64\pm0.10$ & $5.08_{-0.14}^{+0.11}$ & $5.44\pm0.15$ & $5.13_{-0.16}^{+0.13}$ \\ [0.5ex]
\texttt{diskbb} & $kT_{\rm in}$ (keV) & $0.78_{-0.07}^{+0.14}$ & $0.86_{-0.10}^{+0.20}$ & $0.94_{-0.06}^{+0.11}$ & $0.85_{-0.06}^{+0.11}$ & $0.69_{-0.04}^{+0.06}$ & $0.72_{-0.05}^{+0.09}$ \\[0.5ex]
& norm & $1200\pm600$ & $800_{-400}^{+500}$ & $450\pm150$ & $700\pm300$ & $2000_{-700}^{+800}$ & $1400\pm600$ \\ [0.5ex]
& $R_{\rm in}$ (km)$^\dagger$ & $60\pm30$ & $50_{-20}^{+30}$ & $35\pm12$ & $44\pm19$ & $70\pm30$ & $62\pm30$ \\[0.5ex]
\texttt{bbodyrad} & $kT_{\rm bb}$ (keV) & $1.44_{-0.04}^{+0.09}$ & $1.59_{-0.09}^{+0.22}$ & $1.93_{-0.10}^{+0.20}$ & $1.67_{-0.06}^{+0.11}$ & $1.50_{-0.03}^{+0.04}$ & $1.39_{-0.03}^{+0.05}$ \\[0.5ex]
& norm & $220_{-60}^{+40}$ & $130_{-60}^{+40}$ & $49_{-16}^{+12}$ & $100_{-30}^{+20}$ & $220_{-30}^{+20}$ & $230_{-40}^{+30}$ \\[0.5ex]
& $R_{\rm bb}$ (km)$^\dagger$ & $10_{-3}^{+2}$ & $8_{-4}^{+2}$ & $4.9_{-1.6}^{+1.2}$ & $7.0_{-2.0}^{+1.4}$ & $10.4\pm1.0$ &
$10.6_{-1.8}^{+1.4}$ \\[0.5ex] \hline
$\chi^2$/d.o.f. & & 691/621 & 482/478 & 615/633 & 770/641 & 689/656 & 714/632 \\\hline
\multicolumn{2}{r}{Flux$_{2-8\text{ keV}}$ ($10^{-9}$\,erg\,$\textrm{s}^{-1} \textrm{cm}^{-2}$)} & 7.9 & 6.9 & 5.6 & 6.4 & 8.2 & 6.9 \\
\hline
\end{tabular}
\tablefoot{$^\dagger$The inner radius for the \texttt{diskbb} component is estimated assuming an inclination at 65\degr\ and a distance of 7\,kpc \citep{Bandyo99}. For $R_{\rm bb}$ in the \texttt{bbodyrad} component, the same value for the distance was used.}
\end{table*}
\nustar data were used in the energy band 3--30\,keV because at higher energies the background begins to be comparable to or higher than the source count rate. This wide energy range allows for a more detailed characterization of the hard Comptonized component; therefore, we adopted the \texttt{comptt} model instead of the simplified \texttt{bbodyrad} used for the \swift-XRT data. On the other hand, the \nustar energy threshold at 3\,keV makes it more difficult to model the \texttt{diskbb} component, which is partially degenerate with the \texttt{tbabs} component; for this reason, we froze the temperature to $0.8$\,keV as obtained by \swift-XRT. In the \texttt{comptt} model the \texttt{redshift} parameter is set at 0, being \gx a galactic source, and the \texttt{approx} was fixed at 3, meaning a spherical Comptonizing medium, such as the BL/SL\footnote{The scenario with a slab Comptonizing medium (\texttt{approx} at 1) was tested, but there is no improvement, and it almost has the same parameters for the continuum, with the exception of $\tau_\mathrm{sphere}{\sim}2\tau_\mathrm{slab}$, as has also been reported in the literature \citep[see, e.g.,][]{Farinelli23, LaMonaca2025}.}. When this model for the continuum is applied, both absorption and emission lines are visible in the residual, and a bad $\chi^2$/dof (1342/975) is obtained. To model these features, we used two \texttt{gabs} and a \texttt{gaussian} in the simple Model~A reported in Table~\ref{tab:spectrum_nustar}. 
\begin{table*}[!ht]
    \centering
    \caption{Best-fit parameters of the \nustar spectral analysis for \gx in the band 3--30\,keV. Values reported within square brackets are fixed. The estimated unabsorbed flux in 2--8\,keV is ${\sim}8.1\times10^{-9}$\,\fluxcgs, corresponding to a luminosity of ${\sim}4.7\times10^{37}$\,\lum\ for a distance to the source of 7\,kpc \citep{Bandyo99}. The errors are reported at 68\%~CL.}
    \label{tab:spectrum_nustar}
    \begin{tabular}{lrcc}
    \hline\hline
    Model & Parameter (units) & Model A & Model B \\ \\ [-1.7ex] \hline
    \texttt{tbabs} & $N_{\rm H}$ ($10^{22}$ cm$^{-2}$) & $3.0_{-1.0}^{+0.8}$ & $2.29_{-0.03}^{+0.04}$\\ [0.5ex]  \hline\\ [-1.7ex]
    \texttt{diskbb} & $kT_{\rm in}$ (keV) & [0.8] & [0.8] \\
                   & norm ($[R_{\rm in}/D_{10}]^2\cos\theta$) & $600\pm200$ & $280_{-5}^{+6}$\\ [0.5ex]\hline \\ [-1.7ex]
    \texttt{comptt} & $T_0$ (keV) & $1.00\pm0.04$ & $1.037_{-0.004}^{+0.003}$\\
                & $kT$ (keV) & $3.00\pm0.04$ & $2.905\pm0.010$\\
                & $\tau$ & $9.2_{-0.2}^{+0.3}$ & $10.37_{-0.06}^{+0.07}$\\[0.5ex]
                & norm & $0.60_{-0.05}^{+0.04}$ & $0.477_{-0.006}^{+0.008}$\\ [0.5ex]\hline\\ [-1.7ex]
    \texttt{Gaussian} & $E$ (keV) & $5.7_{-0.4}^{+0.3}$ & -- \\[0.5ex]
                & $\sigma$ (keV) & $1.5\pm0.2$ & --  \\
                & norm (photon~cm$^{-2}$~s$^{-1}$) & $0.031_{-0.009}^{+0.013}$ & -- \\[0.5ex]
                & Equivalent width (keV) & $0.27\pm0.07$ & -- \\ \hline\\ [-1.7ex]
\texttt{relxillNS} & Emissivity & -- & $2.20_{-0.03}^{+0.04}$ \\[0.5ex]
                & $R_{\rm in}$ ({ISCO})$^\dagger$ & -- & $4.01_{-0.05}^{+0.07}$ \\[0.5ex]
                & Inclination (deg) & -- & $63.9_{-1.2}^{+1.0}$\\[0.5ex]
                & $\log \xi$ & -- & $2.21_{-0.03}^{+0.02}$\\[0.5ex]
                & $A_{\rm Fe}$ & -- & [1] \\
                & \text{norm ($10^{-3}$)} & -- & $11.6_{-0.7}^{+0.9}$\\[0.5ex]
\hline\\ [-1.7ex]
    \texttt{gabs$_1$} & $E$ (keV) & $8.02\pm0.04$ & $8.07\pm0.03$ \\
                & $\sigma$ (eV) & $87_{-44}^{+54}$ & $79\pm3$ \\[0.5ex]
                & strength (eV) & $17_{-3}^{+4}$ & $15.0_{-0.7}^{+0.8}$\\ [0.5ex]\hline\\ [-1.7ex]
    \texttt{gabs$_2$} & $E$ (keV) & $6.96\pm_{-0.01}^{+0.02}$ & $6.964_{-0.015}^{+0.016}$\\[0.5ex]
                & $\sigma$ (eV) & ${<}150$ & $116_{-4}^{+5}$ \\[0.5ex]
                & strength (eV) & $33_{-3}^{+5}$ & $37.6\pm1.1$\\ [0.5ex]
\hline\\ [-1.7ex]
\multicolumn{2}{r}{$\chi^2/\textrm{d.o.f.}$} & 1059/966 & 1068/964 \\ \hline\\ [-1.7ex]
& & \multicolumn{2}{c}{Cross normalization factors} \\ [0.5ex]\hline\\ [-1.7ex]
\multicolumn{2}{r}{$C_{\rm \nustar-A}$} & [1] & [1] \\
\multicolumn{2}{r}{$C_{\rm \nustar-B}$} & $1.0353\pm0.0014$ & $1.0351\pm0.0019$\\      \hline
\end{tabular}
\tablefoot{$^{\dagger}$The inner radius is given in units of the innermost stable circular orbit (ISCO). }
\end{table*}
The absorption lines are due to the presence of DW \citep{DT12, Dai14, XRISM2025, Rogantini2025, DiMarco25b} and the Gaussian is the main feature of a reflection component; more detailed modeling for the wind applied to \nustar data does not provide a strong improvement due to its energy band and resolution \citep{Saavedra23}. Regarding the reflection component, more detailed modeling is possible with \nustar data applying the \texttt{relxillNS} model \citep{Garcia22} instead of a simple Gaussian; hereafter, this model is referred to as Model~B: \texttt{tbabs*(diskbb+comptt+relxillNS)}. In the \texttt{relxillNS} model, we froze the density parameter, $\log N$, at 19 (i.e., its maximum allowed value), the spin of the NS at 0 \citep{Galloway08, Miller_Miller16, Ludlam24}, the outer radius at $1000\,R_{\rm g}$ and the reflection fraction parameter at $-1$ to directly determine the fraction of photons reflected by the disk. We assumed the same temperature for the continuum blackbody in the \texttt{relxillNS} model and for the seed photons in the \texttt{comptt}; this assumption corresponds to an illumination originating from the BL and/or SL. The same emissivity for the inner and the outer disk was assumed (the power law spectral indices were assumed equal, and the break radius was not used). The best-fit values and errors reported in Table~\ref{tab:spectrum_nustar} were obtained from a Markov chain Monte Carlo (MCMC) with 100 walkers, a burn-in of $10^5$, and a chain length of $2\times 10^5$. No improvement in the $\chi^2$/dof is observed between model~A and model~B, confirming the results obtained by \citealt{Saavedra23}. The accretion disk temperature is in agreement with past observations \citep{Bobrikova24b,DiMarco25b,Saavedra23}; the results obtained for the reflection component show a flatter emissivity, which is compatible with either radially extended illumination of the disk from a slab-like BL or a radially extended corona \citep{Ludlam25}. Furthermore, the large $R_\textrm{in}$ value confirms the effect from the bulge obscuring the inner region of the disk, allowing one to observe only the outer region of the disk. Interestingly, the \texttt{gabs} are compatible with Ly$\alpha$ lines of Fe and Ni, while He$\alpha$ lines are not needed; the result is in contrast with the one by \xrism \citep{XRISM2025}, possibly suggesting a more ionized wind.

Furthermore, aiming to characterize spectral variations along the periodic dip, the spectral analysis of \nustar data was performed in three flux bands. The results are reported in Appendix~\ref{sec:flux_spectra}; no significant variations in the parameters are visible, except for an increase in the absorption in the lower flux interval.

\subsection{Spectropolarimetric analysis}

The \nustar data covered only part of the S1 interval. Due to this limitation and the observed variation in polarization between S1 and S2, the spectropolarimetric analysis was performed for S1, which includes the periodic dip. The spectral model parameters were fixed to the ones in Table~\ref{tab:spectrum_nustar} after removing the \texttt{gabs} components, which cannot be observed by \ixpe due to limited energy resolution and effective area. First, we considered model~A for the \ixpe $I$ spectra, obtaining a $\chi^2$/dof=462/436. Then, assuming in the spectropolarimetric analysis an unpolarized Fe fluorescence line \citep{Churazov02, Veledina24}, the resulting polarization is $\textrm{PD}=4.9\%\pm0.2\%$ and $\textrm{PA}=-44.1\degr\pm1.2\degr$ ($\chi^2$/dof=774/737). We also considered a polarization linearly dependent on energy by using the model \texttt{pollin} having parameters $A1$ and $\psi1$, that are the polarization degree and angle at 1\,keV, while $A_{\textrm{slope}}$ and $\psi_{\textrm{slope}}$ are the linear slopes. The obtained result is $A1=0.9\%\pm0.6\%$, $A_{\textrm{slope}}=(1.5\pm0.2)\%$/keV, $\psi1 = -38\degr \pm 4\degr$, and $\psi_{\textrm{slope}}=(-2.0\pm1.0)$\degr/keV ($\chi^2$/dof=713/735). The consistent Akaike Information Criterion (cAIC) yields values of 42.37 for \texttt{polconst} and -14.39 for \texttt{pollin}, indicating that the latter is strongly preferred and confirming that the PD is energy-dependent, whereas the PA slope is significant only at ${\sim}2\sigma$. When independent polarization is assigned to both \texttt{diskbb} and \texttt{comptt} the best-fit result is ($\chi^2$/dof=711/735): $\textrm{PD}_{\texttt{diskbb}} = 12\% \pm 2\%$ at $\textrm{PA}_{\texttt{diskbb}} = 28\degr \pm 5\degr$ and $\textrm{PD}_{\texttt{comptt}} = 7.9\% \pm 0.4\%$ at $\textrm{PA}_{\texttt{comptt}} = -49.0\degr \pm 1.6\degr$. 

In the case of the Model~B, when a different constant polarization is associated with each spectral component, the results we obtained are reported in Table~\ref{tab:modelB_pol}.
\begin{table*}[!htb]
\centering
\caption{Spectropolarimetric analysis of \gx applying the \texttt{polconst} model to each component in Model~B of Table~\ref{tab:spectrum_nustar}. Values reported within square brackets are fixed.}
\label{tab:modelB_pol}
\begin{tabular}{llccc}
\hline \hline
Spectral component & & Model B1 & Model B2 & Model B3\\
\hline
\texttt{diskbb} & PD & ${<}100$\% &  [4\%]  & [4\%] \\ 
 & PA & -- & [28\degr] & [28\degr]\\ 
 \hline
\texttt{comptt}& PD &  $22\%\pm5\%$ & $13.6\%\pm1.3\%$ & [4\%] \\ 
 & PA (\degr) & $42\degr \pm 7 \degr$ & $-52\degr \pm 3 \degr$ & [$-49\degr$]\\
\hline
\texttt{relxillNS}& PD  & $46\pm19$\% & $15\%\pm3\%$ & $26\%\pm0.8\%$ \\ 
 & PA (\degr) & $53\pm13$\degr & $28\degr\pm6\degr$ & $-43.3\degr \pm 0.8\degr$\\
 \hline
\multicolumn{2}{l}{$\chi^2$/dof} & 797/733 & 801/735 & 978/737 \\ \hline
 \end{tabular}
\tablefoot{The errors are reported at 68\%~CL.}
\end{table*}
In particular, the $\chi^2$/dof is worse than in Model~A, and in B1 the \texttt{diskbb} component has an unconstrained polarization. Aiming to constrain the polarization of the \texttt{relxillNS} component, in Model~B2 we froze the PD of the \texttt{diskbb} component at the value expected from \citealt{Chandrasekhar60} and the PA at the one obtained from Model~A (28\degr). As a further attempt, we applied Model~B3, where the \texttt{diskbb} component has polarization as in Model~B2, and the \texttt{comptt} component has a PD at the expected value from \citealt{st85} with a PA fixed at $-49\degr$, as resulting in Model~A. Comparing the results obtained under these different assumptions, we observed a reflection component that aligns either parallel or orthogonal to the Comptonization, highlighting the degeneracy of these two components \citep{LaMonaca24, LaMonaca24gx340, LaMonaca25gx349+2,LaMonaca2025, Liu26}. 

In conclusion, the spectropolarimetric analysis with present data cannot provide conclusive results for the polarization applying Model~B, while for Model~A shows a \texttt{diskbb} component having PA rotated by ${\sim}80\degr$ with respect to the \texttt{comptt} one; moreover, the disk PD is about twice the expected value \citep{Chandrasekhar60}, and also the Comptonized component shows a PD higher than the one expected for a BL \citep{st85} and/or a SL \citep{Bobrikova25}.

\section{Discussion} \label{sec:discussion}

\subsection{Polarization}

\gx appears to be in the NB  during this new observation, as shown in Fig.~\ref{fig:hid_ccd}, and the measured polarization, considering the observation as a whole, is $\textrm{PD} = 4.4\% \pm 0.3\%$ with $\textrm{PA} = -42.0\degr \pm 1.7\degr$. \gx shows a higher polarization than the other Z sources observed by \ixpe in the NB so far. \ixpe observations of other Z sources showed that the polarimetric properties change with the source state, yielding PD${\sim}$4\% in the HB and less than 2\% in NB and FB \citep{LaMonaca24gx340, LaMonaca2025, DiMarco25a}. Furthermore, none of the other Z sources showed a time variability like that observed in \gx; only for \mbox{XTE J1701$-$462} in NB a possible PA variation was observed, but the PD is almost constant at ${\sim}2$\% along the observation \citep{DiMarco25a}. On the other hand, the accretion disk corona source \mbox{2S~0921$-$630} showed an enhancement of the PD in the predicted eclipsing phase and properties compatible with the presence of scattering in an extended accretion disk corona or in the disk wind \citep{TomaruADC}. 

The behavior of the polarization over time, shown in Fig.~\ref{fig:pd_time_dips}, is complex and somewhat similar to that observed during the previous \gx observations. In particular, this new observation, covering the periodic dip, is characterized by a first segment, S1, dominated by a ``dipping'' phase during which the PD increases, reaching a maximum in the center of the periodic dip, while the PA rotates from north to west; after the end of the periodic dip, the PD decreases and PA rotates from north to east, moving in a sort of high-flux state during a second period, S2.  This trend is well visible in Fig.~\ref{fig:qu_time}, and it is compatible with the idea of polarization tracking the passage of a region bulging out of the outer disk where the accretion stream impacts, along the line of sight between the NS and the observer.

The shallow orbital modulation and the presence of dips indicate \gx as a high-inclination source. The erratic dips are likely due to the passage of clumps forming at the outer region of the accretion disk. The observed X-rays during dips point to the presence of a component due to X-ray scattering within an extended accretion disk corona or a disk wind, as confirmed by \citealt{XRISM2025}. An optically thin oblate ADC, or a disk wind, naturally enhances the PD and may also cause a time-dependent rotation of PA \citep{DiMarco25b}. When a clump or the bulge passes in front of the central emitting region, the extended ADC remains largely unocculted, but a fraction of it and the central source are partially obscured, breaking the original geometry. This alters the ratio of direct-to-scattered photons reaching the observer as a function of time, inducing a rotation in the PA and varying the PD. Alternatively, obscuration of the disk wind can produce identical polarimetric effects as it scatters X-rays like an ADC, driving the observed PA rotations and PD enhancements during dip events.

\begin{figure*}[!hbt]
    \centering
    \includegraphics[width=0.4\linewidth]{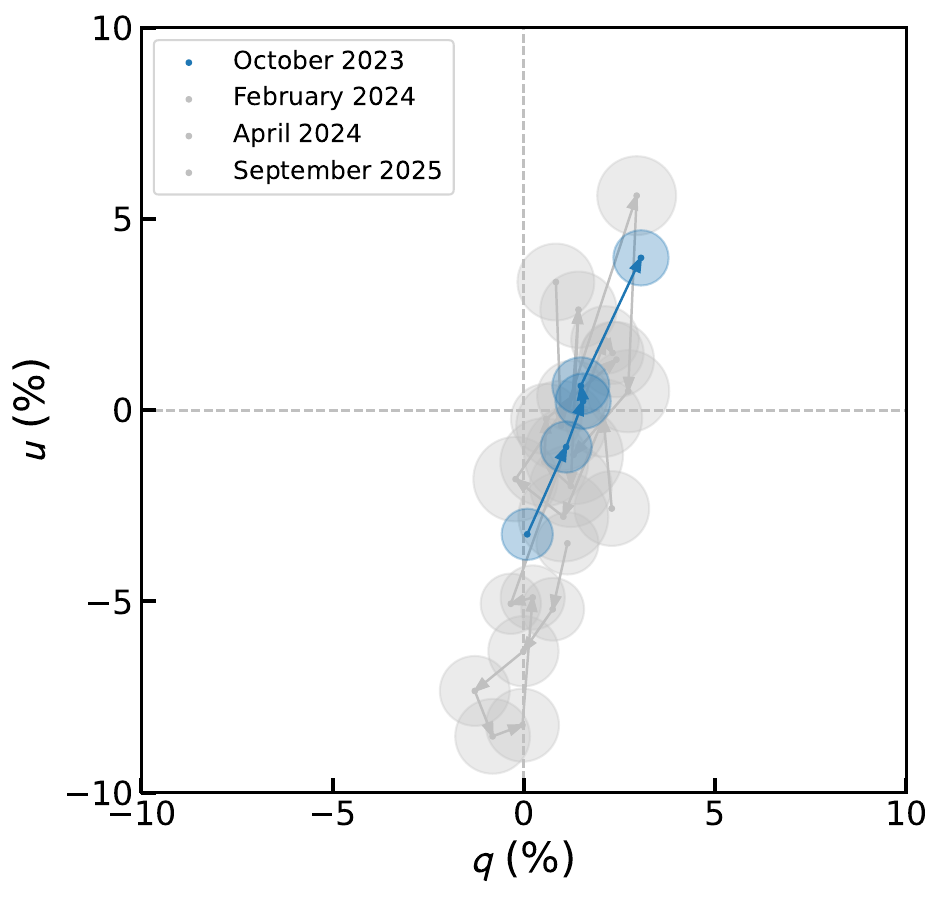}
    \includegraphics[width=0.4\linewidth]{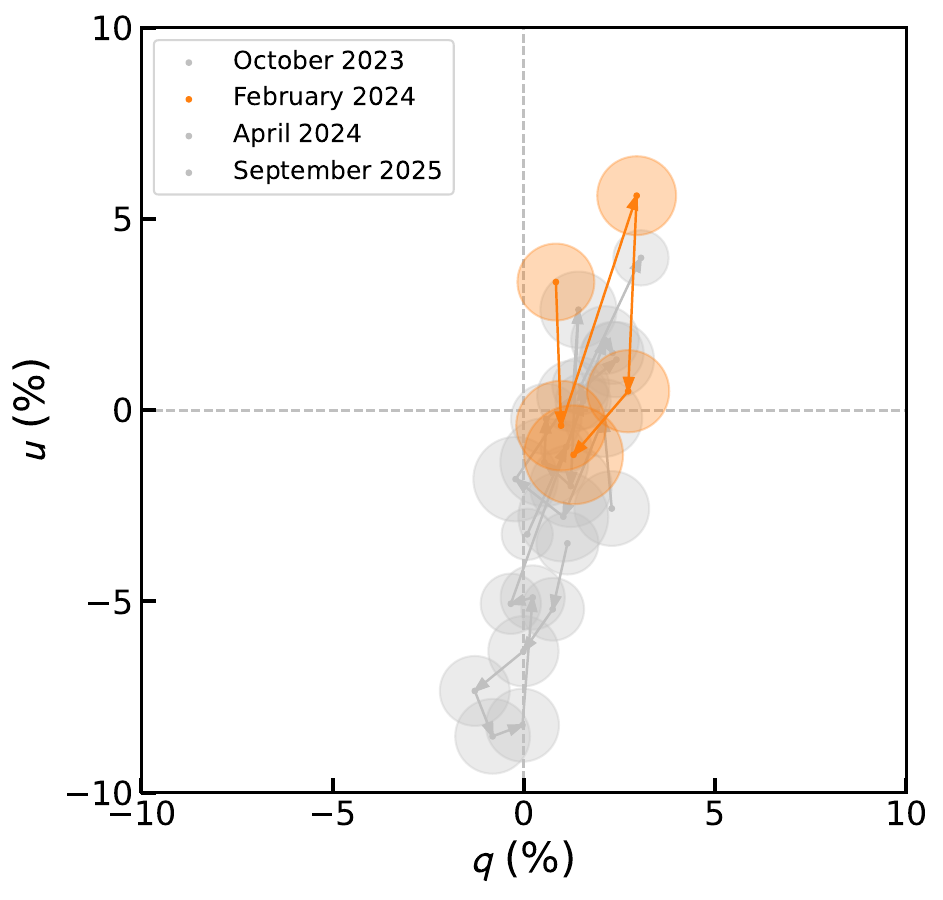}
    \includegraphics[width=0.4\linewidth]{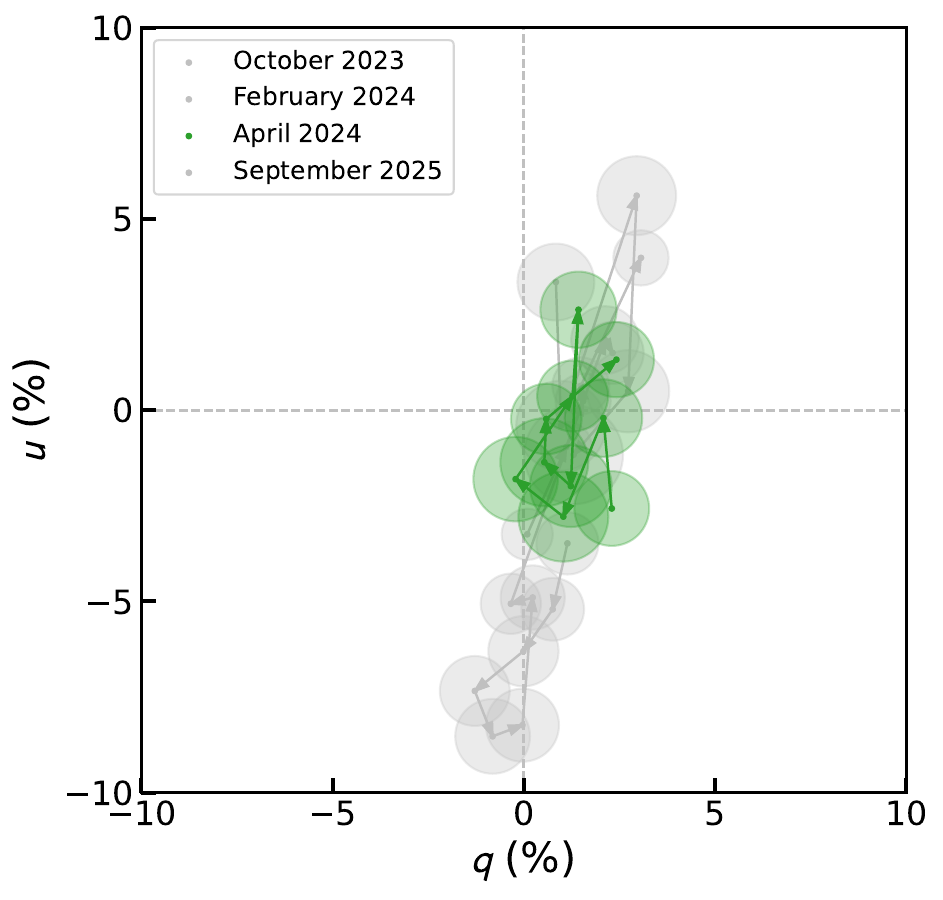}
    \includegraphics[width=0.4\linewidth]{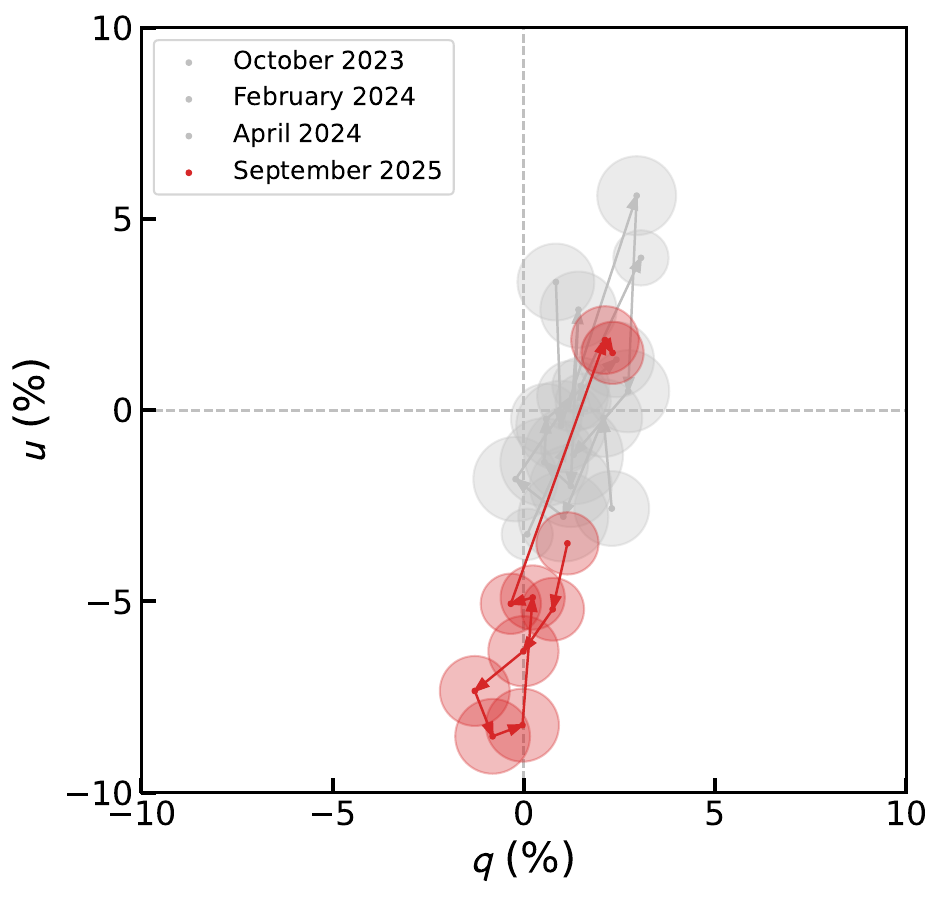}
    \caption{Time dependence of the normalized Stokes parameters $q$ and $u$ in the four \ixpe observations of \gx. The grey circles represent all the available points; the blue points are from the observation of October 2023, applying the same constant time binning as in \citealt{Bobrikova24a}; the orange points are from the observation of February 2024, applying the same constant time binning as in \citealt{Bobrikova24b};  the green points are from the observation of April 2024, applying the same constant time binning as in \citealt{DiMarco25b}; the red points are the same as in Fig.~\ref{fig:qu_time}. The arrows are used to connect each point to the next in time order. Confidence regions are reported at 68\%~CL.}
    \label{fig:previous}   
\end{figure*}

The comparison of all the available observations reported in Fig.~\ref{fig:previous} confirms the idea that PD increases during dips, while the PA rotates. After these dipping events, the polarization returns to its original values, like those during the February 2024 observation, where no dips were observed in the light curve \citep{Bobrikova24b}. 
Furthermore, the polarization as a function of energy shows a linear trend in S1 during the ``dipping'' phase, which is missing in the S2 interval (see Fig.~\ref{fig:pol_energy}). A similar change in the energy dependence was observed in the \gx observation of October 2023 \citep{Bobrikova24a} before and after an erratic dip. As reported in Fig.~\ref{fig:previous}, that observation displayed shifts in the Stokes parameters plane similar to the present ones, confirming the consistency of the results. This variation in the energy dependence may be related to polarization dominated by scattering in S1 and by a more direct emission in S2. In fact, no energy dependence is expected for BL/SL, while, for example, polarization due to scattering in the wind could provide an energy-dependent PD as reported by \cite{TomaruADC}. This result aligns with the idea that during the periodic dip, when the bulge partially obscures the central region, the polarimetric properties of the X-ray emission are dominated by photons due to scattering in the extended accretion disk corona or in the disk wind. 

\subsection{Constraining the geometry of extended accretion disk corona or disk winds from the polarization}

As discussed above, the results reported in Figs.~\ref{fig:qu_time} and Fig.~\ref{fig:pd_time_dips}, agree with an orbital-phase-dependent polarization due to partial obscuration of the X-ray-emitting regions by the passage of the disk bulge along the line of sight. We can use a simplified toy model based on equations by \citealt{BrownMclean1977} to determine the properties of an extended accretion disk corona or an accretion disk wind. The assumption is that the X-rays illuminating the ADC or the DW are unpolarized and emitted from a point-like source. Moreover, we assume that the polarization at the center of the periodic dip is due only to X-rays scattered in the ADC or the DW, having Thomson optical depth \citep{BrownMclean1977}
\begin{equation}
    \tau=\frac{3}{32}\sigma_\textrm{T}\int_{r=0}^{\inf}\int_{\mu=-1}^{1} n(r,\mu) dr d\mu, \label{eq:tau}
\end{equation}
\noindent
where $\sigma_T=6.652\times10^{-25}$\,cm$^{2}$ is the Thomson cross-section, $n(r,\mu)$ the axisymmetric electron density, which depends on the radius $r$, and the polar angle $\mu=\cos \theta$.

\begin{figure}
    \centering
    \includegraphics[width=0.85\linewidth]{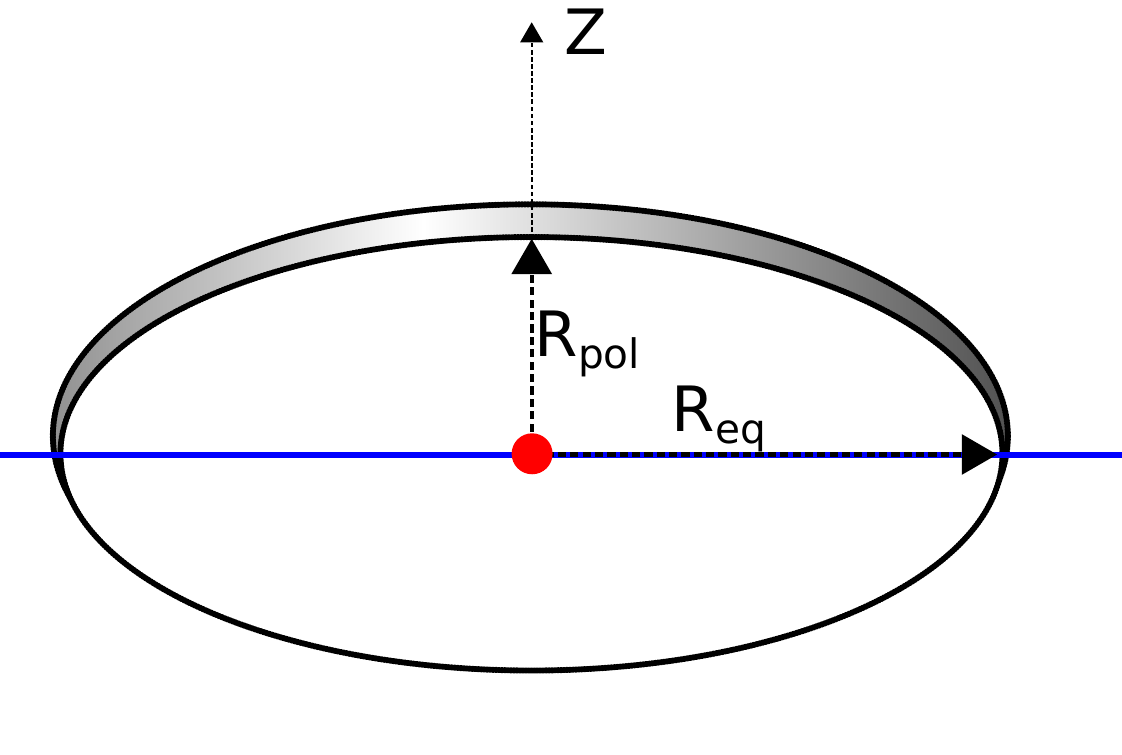}\\
    \includegraphics[width=0.85\linewidth]{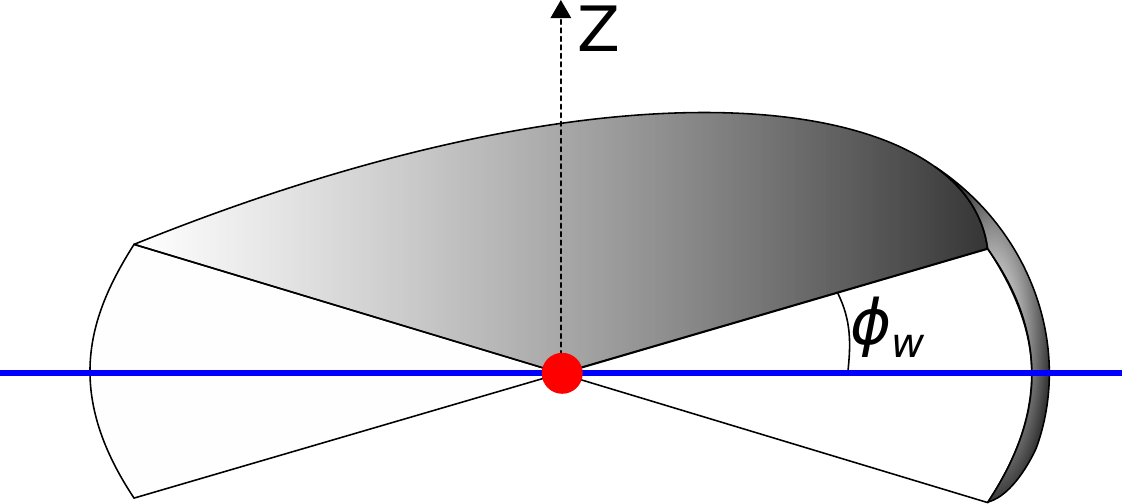}
    \caption{Sketch of the not-to-scale considered geometries. The blue line represents the accretion disk, the red dot the NS, and the shaded grey regions are the external envelope of the extended accretion disk corona (top) and the disk wind (bottom). In the ADC geometry, the main parameters are the equatorial, $R_\textrm{eq}$, and the polar $R_\textrm{pol}$ radii, while for the DW, the main parameter is the opening angle $\phi_\textrm{w}$.}
    \label{fig:geom}
\end{figure}

\begin{figure}[!hbt]
    \centering
    \includegraphics[width=0.75\linewidth]{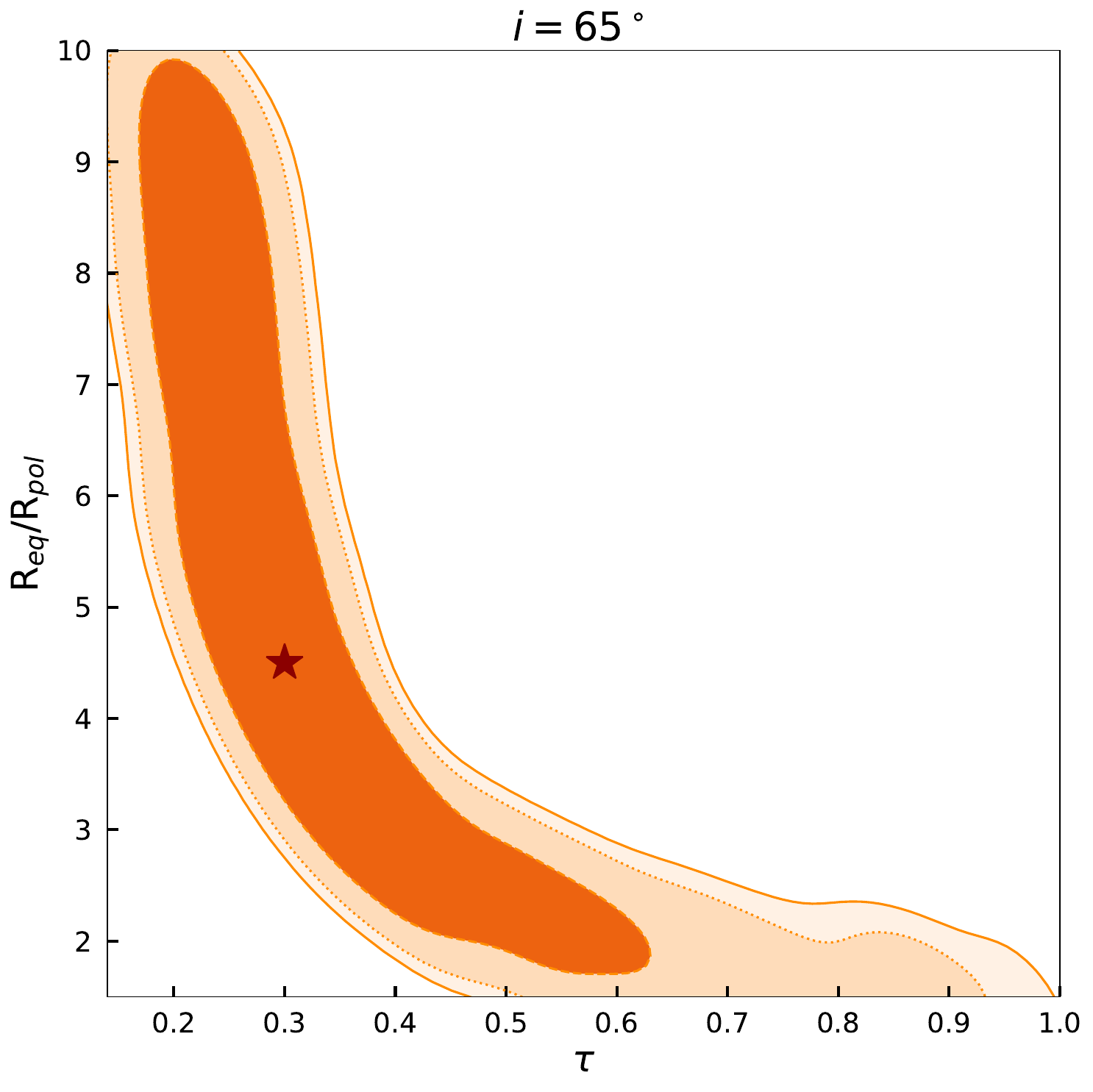}
    \caption{Constraints for the ADC geometry reported in Fig.~\ref{fig:geom}--top, when Eq.~\ref{eq:pd_adc} is applied to the $\textrm{PD}=9.1\% \pm 1.1\%$, as measured at the center of the periodic dip, assuming an inclination of 65\degr. The star marker is the most probable value, while contours show the allowed regions at 68\%, 90\% and 95\% CL, respectively.}
    \label{fig:adc_geom}
\end{figure}

In particular, assuming a scenario with polarization arising from an oblate extended disk corona as in Fig.~\ref{fig:geom}-top, the expected polarization degree is \citep{BrownMclean1977}
\begin{equation}
    \textrm{PD}_\textrm{ADC}=\tau_\textrm{ADC}(1-3\gamma)\sin^2 i,
    \label{eq:pd_adc}
\end{equation}
where $\tau_\textrm{ADC}$ is the optical depth from Eq.~\ref{eq:tau} for this geometry, $i$ is the inclination of the system, and $\gamma$ is a shape factor, which for an oblate geometry is
\begin{equation}
    \gamma=\frac{1}{2(a^2-1)}\left[ \frac{a\sqrt{(a^2-1)}}{\log(a+\sqrt{a^2-1})}-1\right],
\end{equation}
with $a=R_\textrm{eq}/R_\textrm{pol}$, the ratio of the equatorial and polar radii. 
In this geometry, the polarization depends on the source inclination, the parameter $a$, and the optical depth. In Fig.~\ref{fig:adc_geom}, we report the allowed regions for an inclination of 65\degr, in agreement with results obtained in Table~\ref{tab:spectrum_nustar}.
The results are obtained by using a Bayesian analysis in \texttt{ultranest} \citep{ultranest}; the model parameters are determined on the basis of the following likelihood
\begin{equation}
    \mathcal{L}=-0.5\left(\frac{\textrm{PD}_\textrm{model}-\textrm{PD}_\textrm{dip}}{\sigma_{\textrm{PD}_{\textrm{dip}}}}\right)^2
\end{equation}
with $\textrm{PD}_\textrm{dip}=9.1\% \pm 1.1\%$, and the priors having uniform distributions in the following intervals: $\tau=[0;1]$ and $a=R_\textrm{eq}/R_\textrm{pol}=[1;10]$. The results are an equatorial radius at least 1.5 times the polar one and an optical depth ${>}0.14$; the best values are $a=5_{-3}^{+4}$ and $\tau=0.30_{-0.07}^{+0.27}$. 

\begin{figure}
    \centering
    \includegraphics[width=0.75\linewidth]{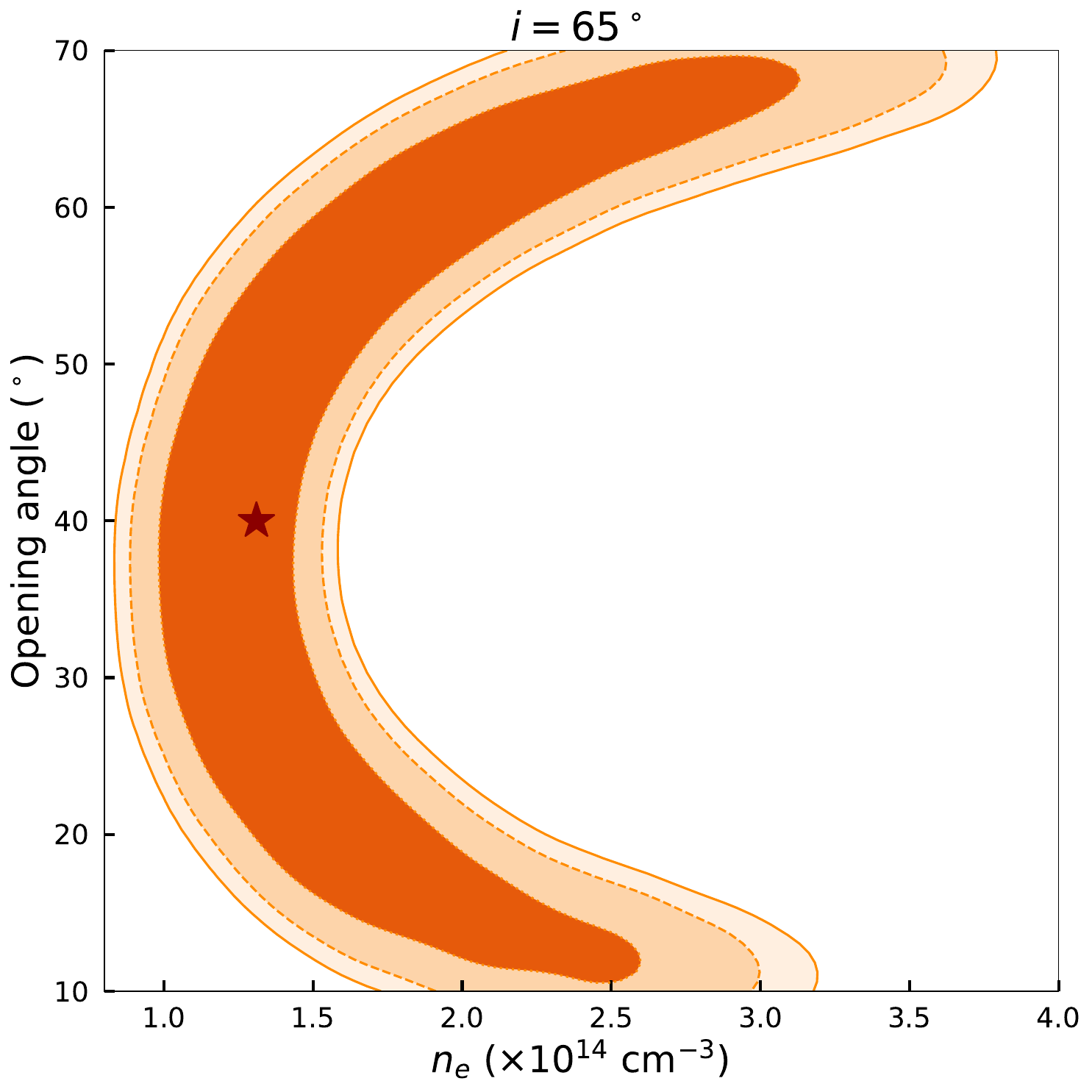}
    \caption{Constraints for the DW geometry reported in Fig.~\ref{fig:geom}--bottom, when Eq.~\ref{eq:pd_dw} is applied to the $\textrm{PD}=9.1\% \pm 1.1\%$, as measured at the center of the periodic dip, assuming an inclination of 65\degr. The star marker is the most probable value, while contours show the allowed regions at 68\%, 90\% and 95\% CL, respectively.}
    \label{fig:dw_geom}
\end{figure}

The scenario of polarization due to scattering in the disk wind can be modeled in a similar way by using the geometry as in Fig.~\ref{fig:geom}-bottom. In this scenario, the expected polarization is \citep{BrownMclean1977}:
\begin{equation}
    \textrm{PD}_\textrm{w}=\frac{3}{16}\sigma_T n_0 R \sin \phi_\textrm{w} \cos^2 \phi_\textrm{w} \sin^2 i ,
    \label{eq:pd_dw}
\end{equation}
with $n_0$ the electron density, $R$ the radius to which the wind extends, $\phi_w$ is the opening angle of the wind (see Fig.~\ref{fig:geom}-bottom), and 
\begin{equation}
    \tau_\textrm{w} = \frac{3}{16}\sigma_T n_0 R \sin \phi_\textrm{w}
\end{equation}
is the Thomson scattering optical depth for this geometry, which is not a constant value, but it depends on $\phi_\textrm{w}$. In this scenario, assuming for the disk wind an outer radius of $1.8\times10^{10}$\,cm \citep{XRISM2025}, inclination 65\degr, and uniform priors $n_0=[10^{12};10^{15}]$\,cm$^{-3}$ and $\phi_w=[10\degr; 70\degr]$ we obtain the result of Fig.~\ref{fig:dw_geom}.
In this geometry, we can determine confidence regions for the electron density and the DW opening angle; the most probable values correspond to $n_0=1.3^{+1.3}_{-0.4}\times10^{14}$\,cm$^{-3}$ and $\phi_w = 40^{\circ}\textrm{}^{+20^\circ}_{-30^\circ}$. These provide an optical depth of $\tau{\sim}$0.23. The results we obtained for both DWs and ADC, albeit not strongly constraining, are in agreement with the values expected by \citealt{DiMarco25b} and the ones obtained by \xrism for the fast and highly ionized wind having density $n_0{\sim}(0.9-1.6)\times10^{14}$\,cm$^{-3}$ \citep{XRISM2025}. 

%Asymmetric winds
Following \citet{nitindala25}, one can model scattering in disk winds that are not axisymmetric (Nitindala et al., in prep). As described above, the bulge is expected to have a higher optical depth than other regions of the disk, breaking the axial symmetry of scattering in accretion disk winds. A higher optical depth implies more scattering and hence a larger PD, similar to the symmetric case \citep[see figure 9 in][]{nitindala25}. On the other hand, in this case, the PA is not simply aligned or orthogonal ($90\degr$ or $0\degr$) to the scattering medium (as in the symmetric scenario) but instead exhibits a rotation, especially across the region where optical depth changes rapidly. The resulting PD and PA depend on the optical depth profile of the wind, the wind opening angle, the central illuminating X-ray source and its intrinsic polarization, and the inclination and azimuthal direction of the observer. While detailed modeling is left to a future paper, preliminary analysis agrees with the scenario presented in \citet{Bobrikova24a} and \citet{DiMarco25b}. The desired PD values shown in Fig.~\ref{fig:pd_time_dips} can be achieved through scattering in the winds. However, the rotation in PA is symmetric across the passage of the disk bulge: the PA rotates towards the main axis of the DWs as we observe it approaching the line of sight, and the PA rotates away from this axis as it moves farther. To better describe the change in PA across the dip (Fig.~\ref{fig:pd_time_dips}), a second polarized component is required in addition to the disk winds that is misaligned with the DW axis.

\subsection{Spectral state}

\begin{figure}[!ht]
    \centering
    \includegraphics[width=0.9\linewidth]{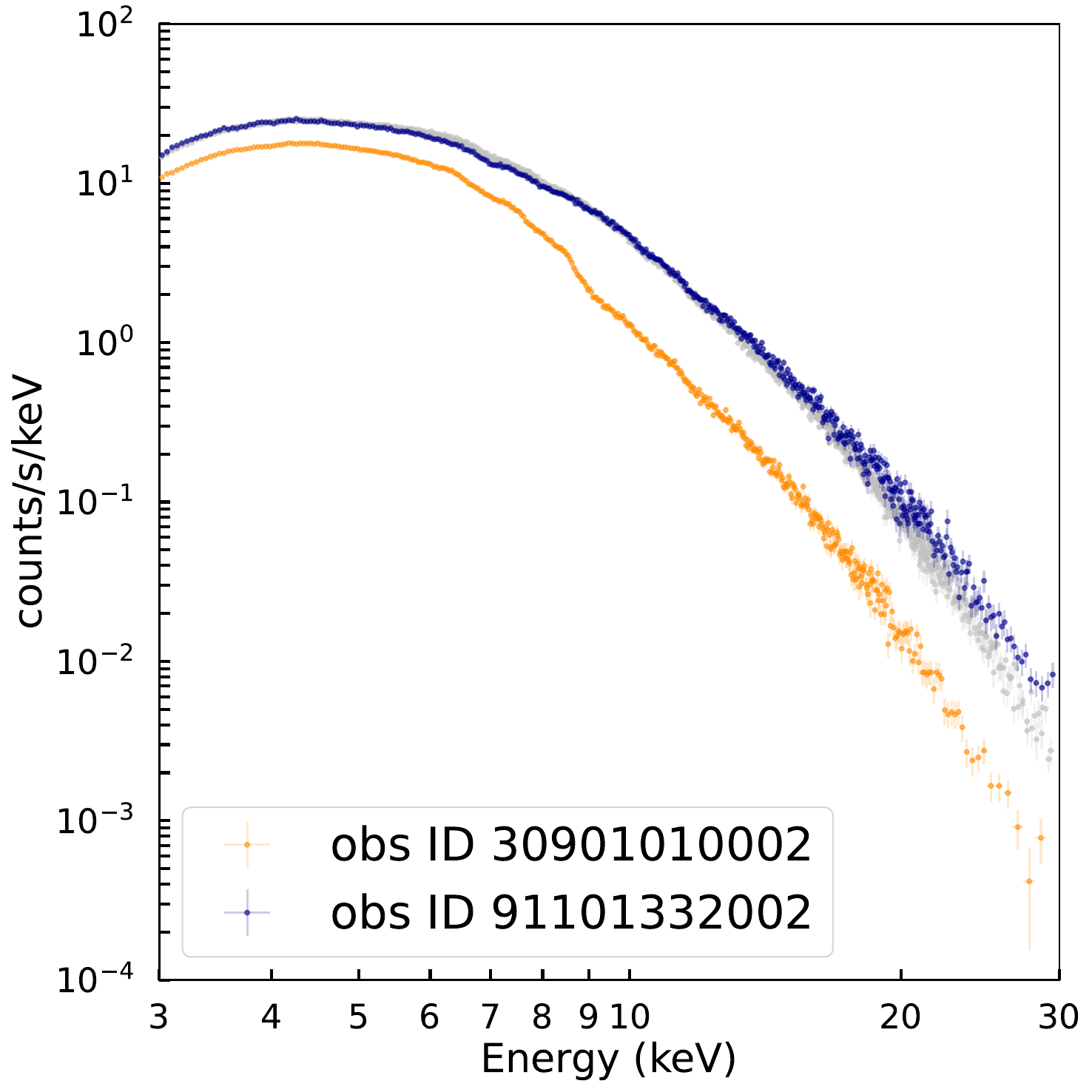}
    \caption{\nustar spectral variability of \gx, with orange points reporting the Observation ID 30901010002 corresponding to a peculiarly absorbed state.}
    \label{fig:nustar_spectra}
\end{figure}

The \nustar data reported in Fig.~\ref{fig:hid_ccd}, and the spectral comparison reported in Fig.~\ref{fig:nustar_spectra}, clearly show the evolution of \gx in a strongly absorbed state, as reported by \cite{XRISM2025}. Despite the presence of the periodic dip, the source had a higher flux and lower absorption in the data presented in our study.
Figure~\ref{fig:hid_ccd} agrees with the Z-type pattern behavior of the source in the CCD diagram as obtained by \citealt{Fridriksson_2015}. On the other hand, the HID is not useful for the \gx classification because of the presence of flux variations due to obscuration and absorption resulting from the inclination of the source. In fact, the dips like the one in the present observation produce a NB crossing the FB, instead of the typical Z shape (see Fig.~\ref{fig:nustar_spectra}); this result confirms the \gx HID diverging from that obtained from other Z sources, making it difficult to disentangle the different branches from it. Furthermore, a similar NB crossing FB behavior was reported by \cite{Stella1985}. Considering that the crossing branch of this new study is the NB during the periodic dip, we used the ephemeris from \citealt{Iaria2014}\footnote{This ephemeris was determined on a dataset spanning from 1996 up to 2013, and no need to add a derivative term was reported.} to determine the orbital phase of the observation analyzed by \cite{Stella1985} which was performed by EXOSAT in 1983 between September 22 at 20:13 UTC and September 23 at 00:18 UTC. The crossing branch was observed at the end of the observation, where a flux decrease with increasing hardness was also reported; assuming the end of the observation as the minimum in the observed flux, the result is an orbital phase of 0.93, which is about $-1.7$\,days with respect to the expected epoch for the periodic dip. The observation performed in 1983 does not show the exit from the dip, and the center of the dip could occur hours after the end of the observation, consistent with the idea that the crossing branch corresponds to the periodic dip of \gx. 

\subsection{Model-dependent polarimetric studies}

The spectral analysis reported in Table~\ref{tab:spectrum1} for the swift data provides a disk temperature ${\sim}0.8$ keV; the \nustar data provided two different broadband models reported in Table~\ref{tab:spectrum_nustar}. Model~B, where the \texttt{relxillNS} model is adopted, yields parameters consistent with partial obscuration of the central region: an inclination angle of ${\sim}$65\degr, an apparent inner disk radius of $R_{in} = 4 R_{isco}$, a flattened emissivity index of $q = 2.2$, and a moderate ionization parameter of $\log\xi = 2.21$. The obtained inclination is compatible with expectations for dipping NS-LMXBs and with other results reported in the literature \citep[see, e.g.,][]{Saavedra23}. During the observed dips, the bulge acts as a geometric filter that fundamentally alters the observed reflection spectrum. Because of that, these parameters describe a geometric line-of-sight effect due to the passage of the bulge in the periodic dip, rather than a truncated accretion disk. The fit results for the inner radius and the emissivity index are direct consequences of this localized obscuration. The ionization parameter of $\log\xi = 2.21$ further corroborates this scenario; the moderate ionization observed during the dip confirms that the primary reflection spectrum originates from material situated farther from the central neutron star. Moreover, the \nustar spectral analysis has been performed for different flux intervals, as reported in Appendix~\ref{sec:flux_spectra} (see Tables~\ref{tab:spectrum_nustar_fluxA} and \ref{tab:spectrum_nustar_fluxB}). 

When the polarization is accounted for in the S1 interval that is covered by \nustar and includes the dip, the polarimetric model that provides the best fit for the data is a linear polarization with PD having a slope of $(1.5\pm0.2)\%$/keV and PA of $(-2.0\pm1.0)$\degr/keV, even if the PA variation is not exactly linear (see Fig.~\ref{fig:pol_energy}). When independent polarization is assigned to both \texttt{diskbb} and \texttt{comptt} in Model~A and the Gaussian is assumed unpolarized, a $\chi^2$/dof in line with the linear trend is obtained for the following polarizations: $\textrm{PD}_{\texttt{diskbb}} = 12\% \pm 2\%$ at $\textrm{PA}_{\texttt{diskbb}} = 28\degr \pm 5\degr$ and $\textrm{PD}_{\texttt{comptt}} = 7.9\% \pm 0.4\%$ at $\textrm{PA}_{\texttt{comptt}} = -49.0\degr \pm 1.6\degr$. In the case of Model~B, inconclusive results are obtained (see Table~\ref{tab:modelB_pol}), the same happens for Model~A applied to different flux intervals (see Appendix~\ref{sec:flux_spectra}). The polarization vectors of the soft and the hard components of the continuum, in systems with aligned BL/SL and disk symmetry axes, are expected to be ${\sim}90\degr$ apart or parallel, depending on the geometry and optical depth of the boundary/spreading layer represented by the Comptonized spectrum \citep{st85}. In this new study, we observe that the \texttt{diskbb} component has a PA  $77\pm5$\degr away from the PA of the \texttt{comptt} component, which is  ${\sim}2.6\sigma$ from a difference in PA of 90\degr; even if the PA is not orthogonal, particular geometries can provide deviations by up to 30\degr\citep{Bobrikova25}. 

\section{Conclusion} \label{sec:conclusion}

We presented the first coordinated IXPE, NuSTAR, and Swift-XRT observations of the highly inclined NS-LMXB \gx, specifically targeted to observe its recurrent 24.5\,day periodic dip. During this campaign, the source was observed in the normal branch, exhibiting a peculiar crossing track in the HID that strongly aligns with historical observations taken during periodic obscuration events.
Our model-independent polarimetric analysis reveals a highly complex and significant temporal correlation between the X-ray polarization properties and the flux variations of the source. As the source enters the periodic dip, caused by the bulge passing in front of the line of sight, the polarization degree increases significantly, reaching a maximum of $9.1\%\pm 1.1\%$ at the center of the dip. Furthermore, we detected a striking ${\sim}60$\degr~rotation in the polarization angle passing from the dip state to the subsequent off-dip state. The energy dependence of the polarization also shifts across these intervals: within the dip, the polarization degree increases linearly with energy, whereas in the off-dip state, it remains constant. This fundamental change supports the hypothesis that the emission during the periodic dip is much affected by scattered photons, while the off-dip state is dominated by direct X-ray emission.
By utilizing a simplified scattering model, we leveraged the polarization maximum at the bottom of the dip to constrain the geometry of the surrounding medium. Assuming that the scattered polarization arises from an oblate extended accretion disk corona, our analysis indicates an equatorial radius that is at least 1.5 times the polar radius, alongside an optical depth ${\sim}0.3$, as suggested by \citep{DiMarco25b}. Alternatively, if the polarization originates from scattering within the disk winds, the model yields a most probable electron density of ${\sim}1.3\times10^{14}$\,cm$^{-3}$ and a wind opening angle of ${\sim}40\degr$, corresponding to an optical depth of ${\sim}0.23$. These geometrical constraints are in clear agreement with recent high-resolution spectroscopic studies, demonstrating the diagnostic power of X-ray polarimetry in mapping the obscured architectures of accreting compact objects.

\begin{acknowledgements}
This research used data products from GO program 2147 (PI A. Di Marco), provided by the IXPE Team (MSFC, SSDC, INAF, and INFN) and distributed with additional software tools by the High-Energy Astrophysics Science Archive Research Center (HEASARC), at NASA Goddard Space Flight Center (GSFC). The Imaging X-ray Polarimetry Explorer (IXPE) is a joint US and Italian mission.  The authors acknowledge the Swift and NuSTAR teams, whose data were used in this research, for promptly scheduling the observations reported here. 
This work was supported by INAF Research Grant “Polarized X-rays from an accreting millisecond pulsar: a pathway to the equation of state of neutron stars (PULSE-X)” (PI: Papitto). AP was partially supported by Fondazione Cariplo/Cassa Depositi e Prestiti through Grant 2023-2560, “Taking the optical pulse of the quickest spinning Neutron Stars: a pilot exploratory study (SPES)”. FLM, AN, AV, AB and JP acknowledge support from the Research Council of Finland grants 355672 and 372881 and Centre of Excellence in Neutron-Star Physics (grant 374064). AB also acknowledges the support of the European Union’s Horizon Europe research and innovation programme under grant agreement No 101131928, project ACME. GI is supported by a Juan de la Cierva fellowship (JDC2024-053550-I). FX  is supported by National Natural Science Foundation of China (grant No. 12373041 and No. 12422306), and Bagui Scholars Program (XF).
\end{acknowledgements}

\bibliographystyle{yahapj}
\bibliography{biblio}

\begin{appendix}

\section{Data handling for X-ray observatories}\label{sec:dataX}

In this study, \ixpe observations were coordinated with \nustar and \swift-XRT to better constrain the spectral model on a broad energy band. In this section, we briefly report on the data handling and extraction applied to the data from the three observatories.

\subsection{\ixpe}

The Imaging X-ray Polarimetry Explorer (\ixpe) is a NASA mission, in partnership with the Italian Space Agency, launched on 2021 December 9. A detailed description of the observatory and its performance is given in \citealt{Weisskopf2022} and \citealt{Soffitta2021}. \ixpe consists of three identical grazing incidence telescopes, providing imaging and spectral polarimetry over the 2--8\,keV energy band with a time resolution better than 10~$\mu$s. Each telescope comprises an X-ray module of mirror assembly (MMA) and a polarization-sensitive detector unit (DU) equipped with a gas pixel detector (GPD) \citep{Baldini21, DiMarco22b}.

\ixpe observation of \gx had a total exposure time of ${\sim}148$\,ks. Data have been processed following the guide reported by \citealt{DiMarco2026}. Thanks to the images collected by the \ixpe telescopes, the source photons were selected within a circular region having a radius of 90\arcsec\ centered on the position of the source, while the background, following the prescription of \citealt{DiMarco23a} for bright sources, was not subtracted. In light of the prescriptions reported by \citealt{DiMarco2026}, after an anomaly occurred on DU2, the background rejection was applied to the data by using the software \texttt{filter\_background.py} version 3.2.
Model-independent analysis has been performed by using the \ixpe dedicated software \textsc{ixpeobssim} \citep{Baldini22} v. 33. Model-dependent analysis and binned analyses have been performed using \textsc{ftool} released in HEASOFT v. 6.35.2; the spectra have been extracted using \texttt{ixpestartx}\footnote{\href{https://heasarc.gsfc.nasa.gov/docs/ixpe/analysis/contributed/ixpestartx.html}{https://heasarc.gsfc.nasa.gov/docs/ixpe/analysis/\\contributed/ixpestartx.html}}. We used the calibration database released on 2026 June 29. For the spectral analysis, the flux (Stokes parameter $I$) energy spectra of the three DUs have been extracted following the weighted approach described in \cite{Baldini22, DiMarco_2022} and data were binned to have at least 30 counts per energy channel, while in the spectro-polarimetric analysis, we applied a constant energy binning of 120 eV for the energy distributions of Stokes parameters $Q$, and $U$.

\subsection{Swift-XRT}

The Neil Gehrels Swift Observatory \citep{Gehrels2004} carries three instruments to enable the most detailed observations of gamma-ray bursts to date; the X-ray Telescope (XRT) is one of them, based on a sensitive, flexible, autonomous X-ray CCD imaging spectrometer. \swift-XRT coordinated observations with \ixpe are used to monitor possible spectral variations and to determine the temperature of the softer \texttt{diskbb} component. \gx is a bright source, and this implies that \swift-XRT observations were performed in Windowed Timing (WT) mode. 

Data were extracted using \texttt{xselect} v2.5b released with HEASOFT v.~6.35.2. Source and background extractions were performed using the imaging capabilities of \swift-XRT. Given the source counting rate $\ge$100\,counts\,s$^{-1}$, data can be affected by pile-up, as reported in the appendix of \citealt{Romano2006}. To address the possible pile-up, the source has been selected in an annular region having an inner radius of 2 pixels and an outer radius of 22 pixels; coherently, the background has been selected in an annulus containing the same number of pixels (e.g., from 80 to 100 pixels).

\subsection{NuSTAR}

The Nuclear Spectroscopic Telescope Array (\nustar) observatory \citep{Harrison2013} consists of two identical X-ray telescope modules, referred to as FPMA and FPMB, providing broadband X-ray imaging, spectroscopy, and timing in the energy range 3--79\,keV. Its angular resolution is 18\arcsec\ (FWHM), while the spectral resolution is 400\,eV (FWHM) at 10\,keV.

The \nustar data were processed with the standard Data Analysis Software (\texttt{nustardas} 11Mar25 v2.1.5) provided under HEASOFT v6.35.2 with the CALDB version released on 2025 October 06. Being \gx a bright source, the following command has been used: \texttt{statusexpr="STATUS==b0000xxx00xxxx000"}. A circular 50\arcsec\ radius region centered on the locations of \gx has been used to extract the source spectrum, while a circular region outside the source with a radius of 120\arcsec has been chosen for the background. The obtained spectra were grouped to have at least 30 counts per bin.

\section{Flux-dependent spectral analysis}\label{sec:flux_spectra}

\begin{figure*}[!ht]
    \centering
    \includegraphics[width=0.3\linewidth]{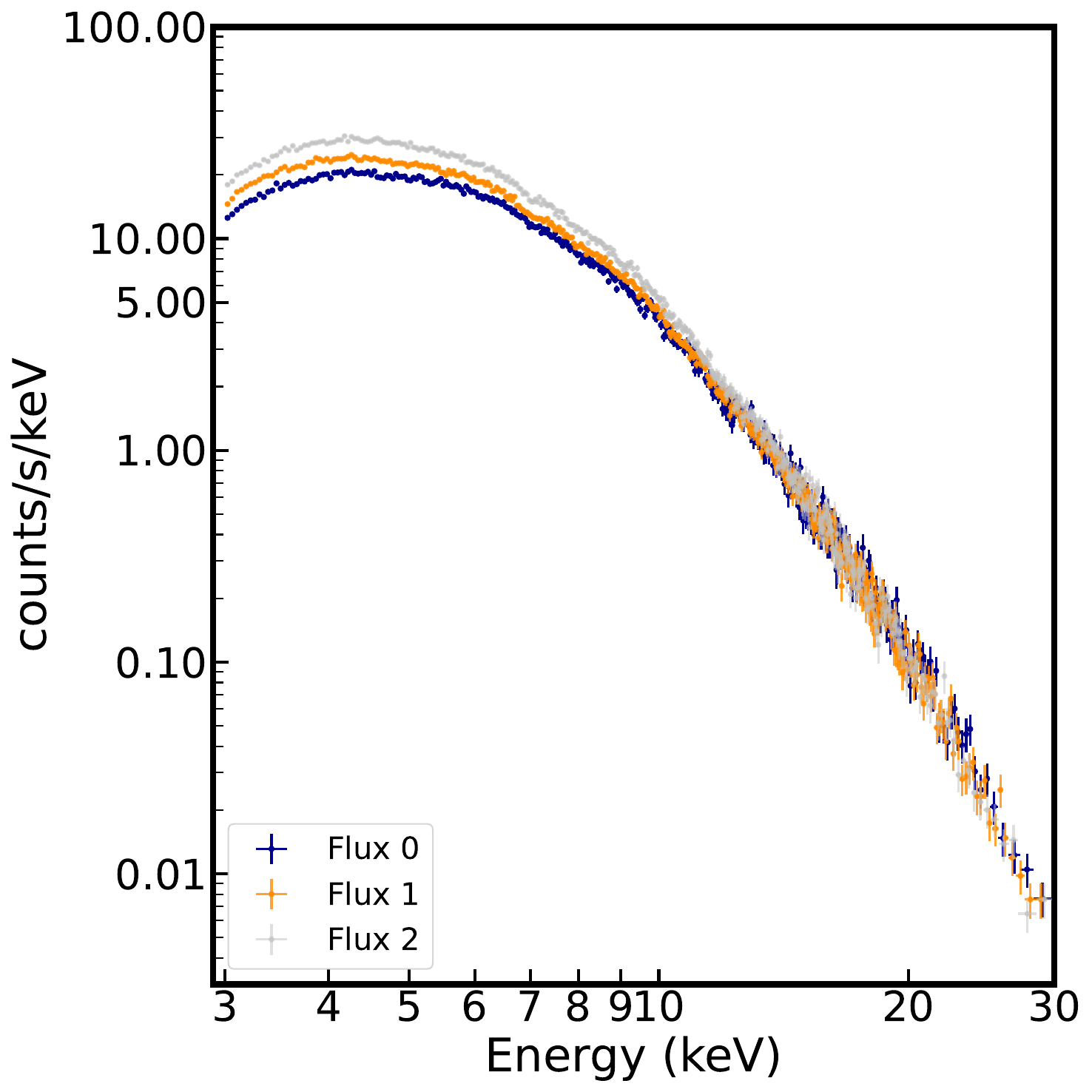}
    \includegraphics[width=0.3\linewidth]{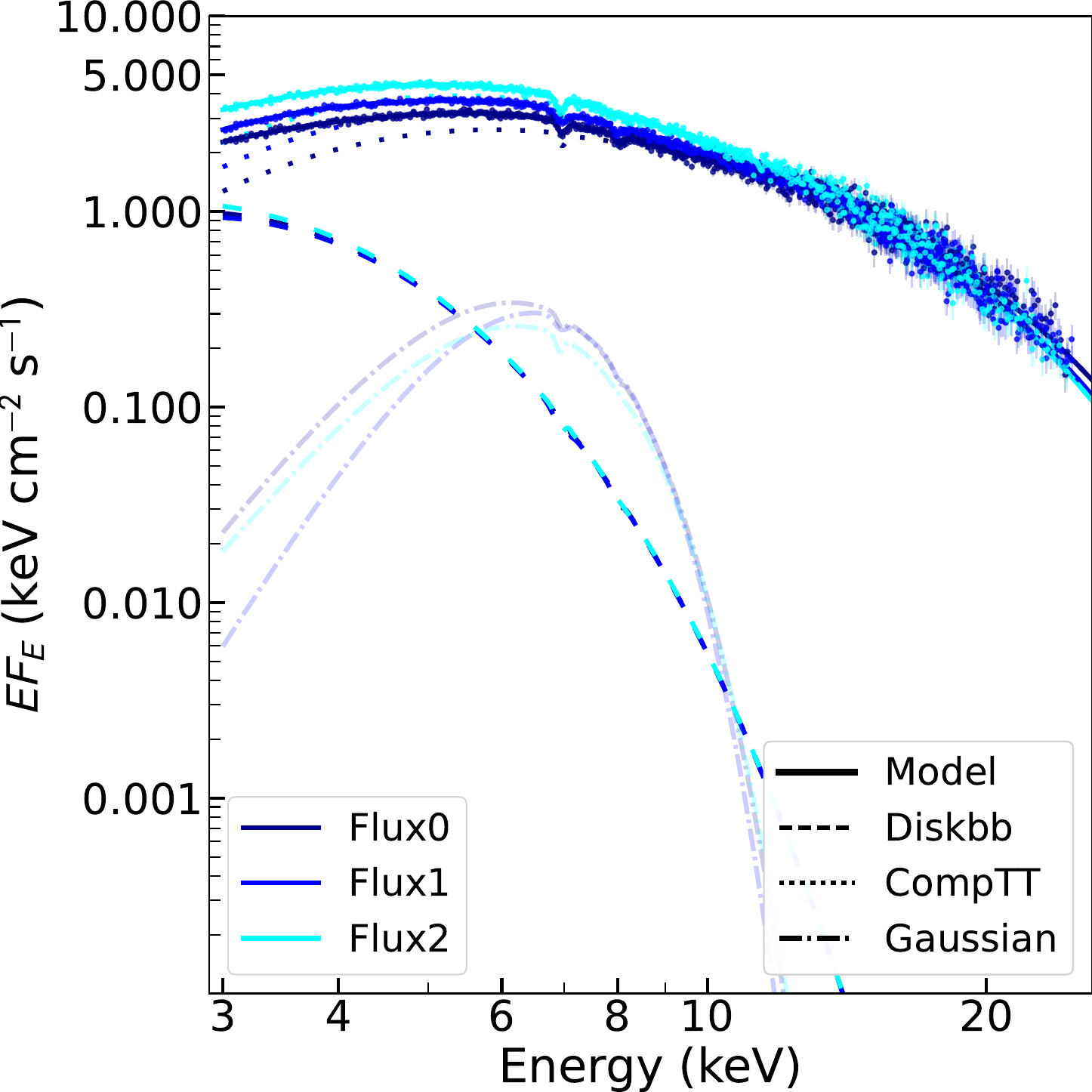}
    \includegraphics[width=0.3\linewidth]{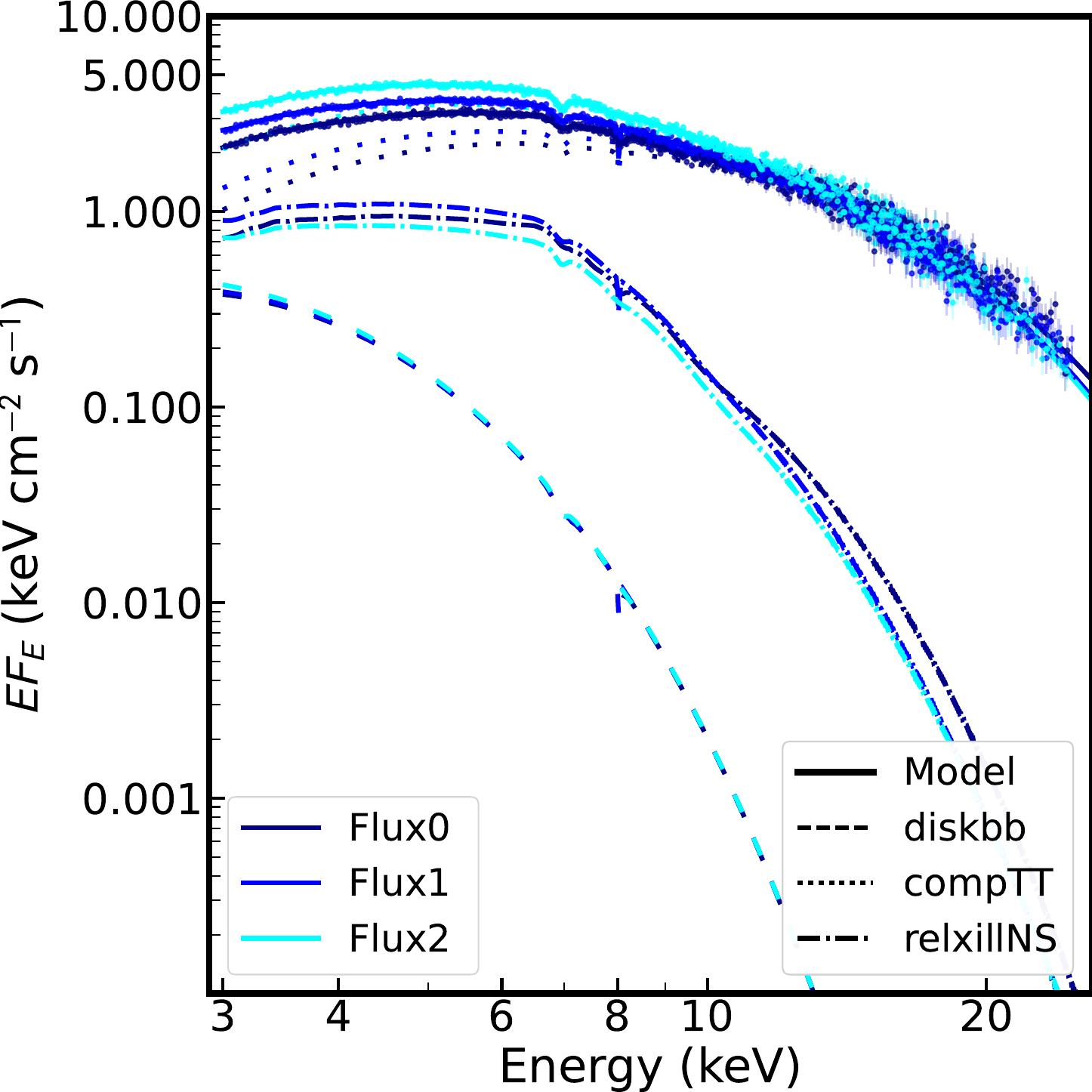}
    \caption{Comparison of \nustar spectra. Left: comparison from \nustar spectra count rates at different flux intervals for the \gx observation performed in September 2025. Center: comparison of the results for Model~A at different fluxes. Right: comparison for Model~B at different fluxes. For both Model~A~and~B the different spectral components are reported as a comparison for each flux interval.}
    \label{fig:nustar_compare_flux}
\end{figure*}

Aiming to characterize spectral variations along the periodic dip, the spectral analysis of \nustar data was performed in three flux bands: flux 0 corresponding to a low flux (${<}105$\,counts/s); flux 1 corresponding to an intermediate flux (between 105 and 120\,counts/s); flux 2 corresponding to a high flux (${>}120$\,counts/s). The difference in the spectrum can be observed in Fig.~\ref{fig:nustar_compare_flux}.

These data have been analyzed using both Model~A and Model~B and the best-fit results are reported in Tables~\ref{tab:spectrum_nustar_fluxA} and \ref{tab:spectrum_nustar_fluxB}, respectively. In this analysis, due to partial degeneracy of parameters for \texttt{tbabs} and \texttt{diskbb}, the latter component has been frozen at values of Table~\ref{tab:spectrum_nustar}, assuming that the reduction in flux is due to variation in the absorption during dips.

\begin{table*}[!htb]
    \centering
        \caption{Best-fit parameters of the \nustar spectral analysis for \gx when different flux intervals are selected and Model~A of Table~\ref{tab:spectrum_nustar} is applied. The errors are reported at 68\%~CL.}
    \label{tab:spectrum_nustar_fluxA}
    \begin{tabular}{crccc}
    \hline \hline
    Model & Parameter (units) & Flux 0 & Flux 1 & Flux 2 \\ \hline \\ [-1.7ex] 
    \texttt{tbabs} & $N_{\rm H}$ ($10^{22}$ cm$^{-2}$) & $2.8\pm0.8$ & $3.3_{-0.5}^{+0.3}$ & $1.9_{-0.9}^{+0.8}$ \\ [0.5ex]\hline \\ [-1.7ex] 
    \texttt{diskbb} & $kT_{\rm in}$ (keV) & \multicolumn{3}{c}{[0.8]} \\
                   & norm ($[R_{\rm in}/D_{10}]^2\cos\theta$) & \multicolumn{3}{c}{[600]} \\ \hline \\ [-1.7ex] 
    \texttt{comptt} & $T_0$ (keV) & $1.05_{-0.04}^{+0.05}$ & $0.98\pm0.03$ & $1.03\pm0.03$ \\ [0.5ex] 
                & $kT$ (keV) & $3.1\pm0.9$ & $2.98_{-0.06}^{+0.05}$ & $2.96\pm0.08$\\
                & $\tau$ & $9.5^{+0.5}_{-0.4}$ & $9.3_{-0.3}^{+0.4}$ & $8.7_{-0.4}^{+0.5}$ \\ [0.5ex] 
                & norm & $0.43_{-0.04}^{+0.03}$ & $0.60_{-0.03}^{+0.02}$ & $0.71_{-0.06}^{+0.04}$\\ [0.5ex] \hline \\ [-1.7ex] 
    \texttt{Gaussian} & $E$ (keV) & $5.4_{-0.6}^{+0.4}$ & $5.9_{-0.3}^{+0.2}$ & $5.4_{-0.6}^{+0.8}$ \\ [0.5ex] 
                & $\sigma$ (keV) & $1.5\pm0.2$ & $1.37_{-0.10}^{+0.18}$ & $1.6_{-0.3}^{+0.4}$ \\ [0.5ex] 
                & norm (photon~cm$^{-2}$~s$^{-1}$) & $0.041_{-0.014}^{+0.029}$ & $0.028_{-0.007}^{+0.013}$ & $0.030_{-0.015}^{+0.055}$\\ [0.5ex] 
                & Equivalent width (keV) & $0.38\pm0.14$ & $0.27\pm0.07$ & $0.20_{-0.18}^{+0.08}$ \\ [0.5ex]  \hline \\ [-1.7ex] 
    \texttt{gabs$_1$} & $E$ (keV) & [8.02] & $8.00_{-0.07}^{+0.05}$ & $8.02_{-0.10}^{+0.09}$ \\ [0.5ex]
                & $\sigma$ (eV) & [79] & ${<}200$ & $190_{-80}^{+100}$\\ [0.5ex]
                & strength (eV) & $17\pm6$ & $20_{-6}^{+10}$ & $20_{-7}^{+10}$\\ [0.5ex] \hline \\ [-1.7ex] 
    \texttt{gabs$_2$} & $E$ (keV) & $6.98_{-0.03}^{+0.04}$ & $6.96\pm0.02$ & $6.94\pm0.03$ \\
                & $\sigma$ (eV) & ${<}160$ & $120\pm40$ & $70\pm60$ \\
                & strength (eV) & $28_{-4}^{+5}$ & $39_{-6}^{+8}$ & $32_{-4}^{+6}$ \\ [0.5ex]
\hline \\ [-1.7ex] 
\multicolumn{2}{r}{$\chi^2/\textrm{d.o.f.}$} & 796/799 & 924/854 & 853/801 \\ \hline \\ [-1.7ex] 
& & \multicolumn{3}{c}{Cross normalization factors} \\\hline \\ [-1.7ex] 
\multicolumn{2}{r}{$C_{\rm \nustar-A}$} & [1] & [1] & [1] \\
\multicolumn{2}{r}{$C_{\rm \nustar-B}$} & $1.037\pm0.003$ & $1.035\pm0.002$ & $1.035\pm0.002$ \\      
\hline \\ [-1.7ex] 
\multicolumn{2}{r}{Flux$_{\textrm{2--8\,keV}}$ ($\times10^{-9}$ \fluxcgs)} & 5.6 & 6.4 & 7.9 \\ \hline
\end{tabular}
\end{table*}

\begin{table*}[!htb]
    \centering
        \caption{Best-fit parameters of the \nustar spectral analysis for \gx when different flux intervals are selected and Model~B of Table~\ref{tab:spectrum_nustar} is applied. The errors are reported at 68\%~CL.}
    \label{tab:spectrum_nustar_fluxB}
    \begin{tabular}{crccc}
    \hline\hline 
    Model & Parameter (units) & Flux 0 & Flux 1 & Flux 2 \\ \hline \\ [-1.7ex] 
    \texttt{tbabs} & $N_{\rm H}$ ($10^{22}$ cm$^{-2}$) & $2.4\pm0.2$ & $1.91_{-0.12}^{+0.15}$ & $1.16_{-0.11}^{+0.18}$ \\[0.5ex] \hline\\ [-1.7ex] 
    \texttt{diskbb} & $kT_{\rm in}$ (keV) & \multicolumn{3}{c}{[0.8]}  \\
                   & norm ($[R_{\rm in}/D_{10}]^2\cos\theta$) & \multicolumn{3}{c}{[220]}  \\ \hline\\ [-1.7ex] 
    \texttt{comptt} & $T_0$ (keV) & $1.05\pm0.02$ & $1.02\pm0.02$ & $1.03\pm0.02$ \\
                & $kT$ (keV) & $3.01_{-0.03}^{+0.04}$ & $2.88\pm0.05$ & $2.89_{-0.07}^{+0.05}$\\[0.5ex]
                & $\tau$ & $10.9_{-0.5}^{+0.4}$ & $10.6_{-0.4}^{+0.5}$ & $9.4_{-0.3}^{+0.4}$ \\[0.5ex]
                & norm & $0.36_{-0.02}^{+0.03}$ & $0.46\pm0.02$ & $0.64_{-0.02}^{+0.04}$\\[0.5ex] \hline\\ [-1.7ex] 
\texttt{relxillNS} & Emissivity & \multicolumn{3}{c}{[2.2]}  \\
                & $R_{\rm in}$ ({ISCO}) & \multicolumn{3}{c}{[4]}  \\
                & $R_{\rm out}$ ($GM/c^2$) & \multicolumn{3}{c}{[1000]}  \\
                & Inclination (deg) & $62\pm3$ & $66_{-3}^{+2}$ & $71_{-2}^{+3}$ \\[0.5ex]
                & $\log \xi$ & $2.19_{-0.12}^{+0.08}$ & $2.15_{-0.10}^{+0.09}$ & $2.17_{-0.21}^{+0.11}$\\[0.5ex]
                & $A_{\rm Fe}$ & \multicolumn{3}{c}{[1]} \\
                & $\log N$ & \multicolumn{3}{c}{[19]}  \\
                & \text{norm ($10^{-3}$)} & $10\pm2$ & $14\pm2$ & $12_{-2}^{+4}$\\[0.5ex]
\hline\\ [-1.7ex] 
    \texttt{gabs$_1$} & $E$ (keV) & [8.07] & $8.03_{-0.06}^{+0.05}$ & $8.04_{-0.04}^{+0.07}$ \\[0.5ex]
                & $\sigma$ (eV) & [79] & $22\pm5$ & $210\pm20$ \\
                & strength (eV) & $16\pm4$ & $17\pm5$ & $20\pm2$\\ \hline\\ [-1.7ex] 
    \texttt{gabs$_2$} & $E$ (keV) & $6.99_{-0.04}^{+0.05}$ & $6.96\pm0.02$ & $6.96\pm0.03$  \\[0.5ex]
                & $\sigma$ (eV) & $130_{-60}^{+70}$ & $124_{-16}^{+12}$ & $100\pm20$ \\[0.5ex]
                & strength (eV) & $34_{-7}^{+10}$ & $41\pm4$ & $35\pm5$ \\ [0.5ex]
\hline\\ [-1.7ex] 
\multicolumn{2}{r}{$\chi^2/\textrm{d.o.f.}$} & 801/799 & 930/854 & 854/801 \\ \hline\\ [-1.7ex] 
& & \multicolumn{3}{c}{Cross normalization factors} \\\hline\\ [-1.7ex] 
\multicolumn{2}{r}{$C_{\rm \nustar-A}$} & [1] & [1] & [1] \\
\multicolumn{2}{r}{$C_{\rm \nustar-B}$} & $1.036\pm0.004$ & $1.035\pm0.002$ 
& $1.034\pm0.003$ \\      
\hline \\ [-1.7ex] 
\multicolumn{2}{r}{Flux$_{\textrm{2--8\,keV}}$ ($\times10^{-9}$ \fluxcgs)} & 5.5 & 6.5 & 7.9 \\ 
\multicolumn{2}{r}{Flux$_{\textrm{disk}}$/Flux$_{\textrm{tot}}$}  & 0.123 & 0.109 & 0.098 \\ 
\multicolumn{2}{r}{Flux$_{\textrm{comp}}$/Flux$_{\textrm{tot}}$}  & 0.544 & 0.578 & 0.691 \\ 
\multicolumn{2}{r}{Flux$_{\textrm{ref}}$/Flux$_{\textrm{tot}}$}  & 0.333 & 0.313 & 0.211 \\ \hline
\end{tabular}
\end{table*}

As a last study, \ixpe data in the S1 time interval have been divided into 3 fluxes: Flux 0 (${<}28$\,cps), Flux 1 ($28$--$34$\,cps) and Flux 1 (${>}34$\,cps). Considering that spectropolarimetric analysis for Model~B results in unconstrained polarization for the different components, this spectropolarimetric analysis is performed only in terms of Model~A using the best-fit parameters in each flux interval from Table~\ref{tab:spectrum_nustar_fluxA}. When only the Gaussian is assumed unpolarized, the resulting polarization degree associated with the \texttt{diskbb+comptt} component is $7.4\%\pm0.6\%$, $4.0\%\pm0.3\%$, and $3.1\%\pm0.3\%$ for Flux 0, Flux1 and Flux 2, respectively. It shows an indication of a decrease in the PD as the flux increases. Similarly, the following PAs are obtained: $-47\degr \pm 2 \degr$ for Flux 0; $-42\degr \pm 2 \degr$ for Flux 1; $-35\degr \pm 3 \degr$ for Flux 2. The PA shows a possible indication of rotation, with the increasing flux confirming the results obtained by \citealt{DiMarco25b}. The polarization associated with each spectral component has also been investigated, and the results of this study are reported in Table~\ref{tab:modelA_pol_flux}.

\begin{table*}[!hbt]
\centering
\caption{Weighted spectropolarimetric analysis of \gx applying the \texttt{polconst} model to each component in Model~A in three different flux intervals.}
\label{tab:modelA_pol_flux}
\begin{tabular}{llccc}
 \hline\hline
Spectral component & & Flux 0 & Flux 1 & Flux 2 \\ \hline
\texttt{diskbb} & PD & ${<10}\%$ &  $12\%\pm3\%$  & $24\%\pm6\%$ \\ 
 & PA & -- & $29\degr \pm 8 \degr$ & $30\degr \pm 7 \degr$\\ 
 \hline
\texttt{comptt}& PD &  $8.4\%\pm1.2\%$ & $6.9\%\pm0.7\%$ & $5.1\%\pm0.6\%$ \\ 
 & PA (\degr) & $-48\degr \pm 4 \degr$ & $-48\degr \pm 3 \degr$ & $-45\degr \pm 3 \degr$ \\
\hline
\multicolumn{2}{l}{$\chi^2$/dof} & 621/707 & 659/732 & 733/733 \\ \hline
 \end{tabular}
\tablefoot{The errors are reported at 68\%~CL.\\}
\end{table*}

\end{appendix}

\clearpage

\label{LastPage} 
%-------------------------------------------------------------

\end{document}